\documentclass[a4paper,fleqn,usenatbib,useAMS]{mnras}
\usepackage{graphics}
\usepackage{graphicx}
\usepackage{amsmath}    
\usepackage{amssymb}    
\usepackage{multicol}        
\usepackage{bm}     
\usepackage{pdflscape}  
\usepackage{xcolor}
\usepackage{ae,aecompl}
\usepackage{subfigure}
\usepackage[english]{babel}
\usepackage{times}
\usepackage{physics}
\usepackage{etoolbox}
\usepackage{paralist}
\usepackage[inline]{enumitem}
\newlist{mycompactenum}{enumerate}{1}
\setlist[mycompactenum,1]{nosep,label=\arabic*.}
\usepackage{paralist}
\usepackage{etoolbox}
\makeatletter
\patchcmd\@combinedblfloats{\box\@outputbox}{\unvbox\@outputbox}{}{%
    \errmessage{\noexpand\@combinedblfloats could not be patched}%
}%
\makeatother
\newcommand{\wfirst}{\textit{RST}}

\title[Detecting discs in microlensing with \wfirst]{Detecting the inner regions of discs around sources of microlensing with \textit{Roman Space Telescope}}
\author[Sajadian et al.]{Sedighe~Sajadian$~^{1}$\thanks{E-mail: s.sajadian@iut.ac.ir}, Ali~Salehi$~^{1}$\\
$^{1}$Department~of~Physics,~Isfahan~University~of~Technology,~Isfahan~84156-83111,~Iran}
\date{Accepted XXX. Received YYY; in original form ZZZ}
\pubyear{2020}

\begin{document}
\maketitle
\begin{abstract}
	
The inner region of circumstellar discs makes an extra near-infrared emission (NIR bump). Detecting and studying these NIR bumps from nearby stars have been done mostly through infrared interferometry. In this work, we study the feasibility of detecting NIR bumps for Galactic bulge stars through microlensing from observations by \textit{The Nancy Grace Roman Space Telescope} (\wfirst) survey. We first simulate microlensing light curves from source stars with discs in near-infrared. Four main conclusions can be extracted from the simulations. (i) If the lens is crossing the disc inner radius, two extra and wide peaks appear and the main peak of microlensing light curve is flattened. (ii) In microlensing events with the lens impact parameters larger than the disc inner radius, the disc can break the symmetry of light curves with respect to the time of closest approach. (iii) In caustic-crossing binary microlensing, the discs produce wide peaks right before entering and immediately after exiting from the caustic curves. (iv) The disc-induced perturbations are larger in the W149 filter than in the Z087 filter, unless the lens crosses the disc condensation radius. By performing a Monte Carlo simulation, the probabilities of detecting the disc perturbations by \wfirst~are estimated $\sim 3$ and $20$ per cent in single and binary microlensing, respectively. We anticipate that \wfirst~detects around $109$ disc-induced perturbations during its microlensing survey if $5$ per cent of its source stars have discs. 

\end{abstract}

\begin{keywords}
gravitational lensing: micro,~(stars:) circumstellar matter,~space vehicles,~methods: numerical	
\end{keywords}

\section{Introduction}
As a star evolves from its birth, the remnants of the initial protostellar accretion disc form a protoplanetary disc. These discs are flattened and mostly contain cool dust and gas. These discs are thick in the beginning, and they become thinner as the planetary systems develop. Some of the reasons are (i) accreting onto their host stars, (ii) making planetesimals, asteroids or even planets due to viscosity of the grains (iii) evaporation processes through outflows. These protoplanetary discs generally survive for several million years \citep[see, e.g.,][]{williams2011}. Hence, discs provide a natural laboratory to study the birth of solar systems and the exoplanet formation.

Discs around stars generally consist of two regions, inner and outer. These two meet at the location with the effective temperature around $1500~$K, at which the condensation of dust grains happens. So the inner region (the inner disc) contains pure gas and the outer region (the outer disc) includes cool dust and gas. The inner disc extends to around $0.1-1~$au from the host star. This part of the protoplanetary discs emits thermal radiation in near-infrared, which can be resolved through near-infrared (NIR) interferometry \citep[see, e.g., ][]{dullemond2010,Hoadley2015}. This NIR bump is very strong for Herbig Ae/Be stars, whereas it is weaker for T Tauri stars which are brighter in NIR \citep{klarmann2017,Eisner2010}. Generally, these discs likely exist around low mass stars rather than massive ones. The near-infrared excess (NIR bump) emissions from the inner disc decrease with the stellar age \citep{Haisch2001}. The outer disces can also be detected through their millimeter and infrared (thermal) and partially visible (reflected) emissions \citep{Pinte2008,Throop2001,red2019}. This wide wavelength range of disc emission is due to the high variation of the temperature throughout discs.

Observations of discs in infrared was started by \textit{The Infrared Astronomical Satellite} (\textit{IRAS}) \citep{Strom1989}. The NIR emission originates from a region of one au, and thus it can be resolved for objects located within $\sim100~$pc far from us with a resolution of $10~$milliarcsec. In order to obtain more information about the inner regions of circumstellar discs by spatially resolving them or detecting discs around low mass stars, higher-resolution observations by IR interferometers are needed \citep{Sargent1987}. Examples of such instruments are \textit{Palomar Testbed Interferometer} (\textit{PTI}) \citep{Malbet1998}, \textit{Infrared Optical Telescope Array} (\textit{IOTA}) interferometer \citep{Millan1999} and \textit{Spitzer Space Telescope} \citep{Muzerolle2009}. In 1994, \textit{Hubble Space Telescope} (\textit{HST}), obtained multiple pictures of discs orbiting stars seen in silhouette against the bright background of the Orion nebula \citep{Beckwith2000}.

Although NIR interferometry is the best method for detecting and studying the inner discs through their NIR bumps, it is limited to bright, very young and close stellar objects. For this reason, all confirmed discs are closer than $3500~$pc \footnote{\url{http://www.circumstellardisks.org/}}. The thermal emission from the inner discs around bulge stars is too weak to be detected through NIR interferometry.

In this work, we evaluate the ability of the \wfirst~survey to discern these emissions in its microlensing observations. In this regard, we mention three key points. (i) \wfirst~microlensing observations will be conducted in the W149 filter. The mean wavelength of this filter, $1.49~\mu$m, coincides with the wavelength of NIR bump emission from the inner disc. (ii) Source stars of microlensing events detectable from the \wfirst~survey will be, on average, much fainter and less massive than the source of the events detectable in the current microlensing surveys \citep{Penny2019}. The discs around less massive stars survive longer \citep{Pascucci2011}. Hence, the probability for the less massive stars to have discs is higher. (iii) \wfirst~will probe microlensing events towards the Galactic bulge. Although the majority of Galactic bulge stars are old, $\sim 10-20$ per cent of them are younger than $5~$Gyrs \citep{Pfuhl2011}. In order to show the ability of \wfirst~to detect of these weak signals, we do a Monte Carlo simulation and estimate the probability of detecting these discs through microlensing observations.

In the following section, the models for thermal and reflected Spectral Energy Distribution (SED) of discs used in this paper will be introduced. In the section~\ref{three}, we discuss the microlensing phenomenology of lensing events occurring on source stars with discs. We classify disc-induced perturbations in lensing light curves. In the section~\ref{four}, the probability of detecting NIR bumps in microlensing events which will be detected by \wfirst~is estimated by performing a Monte Carlo simulation. In the last section, we summarize the results and conclusions.

\begin{figure*}
    \centering
    \includegraphics[width=0.49\textwidth]{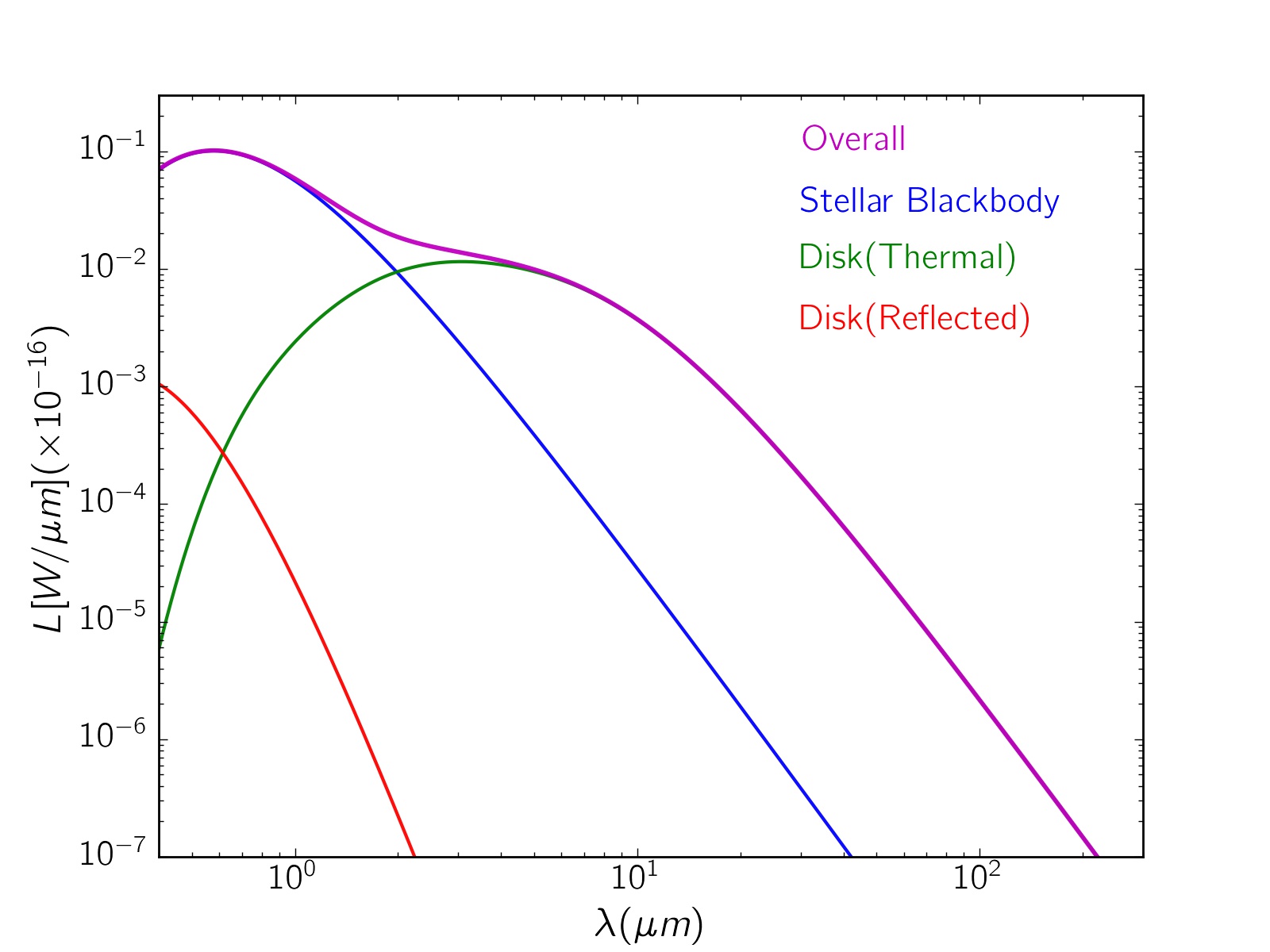}
    \includegraphics[width=0.49\textwidth]{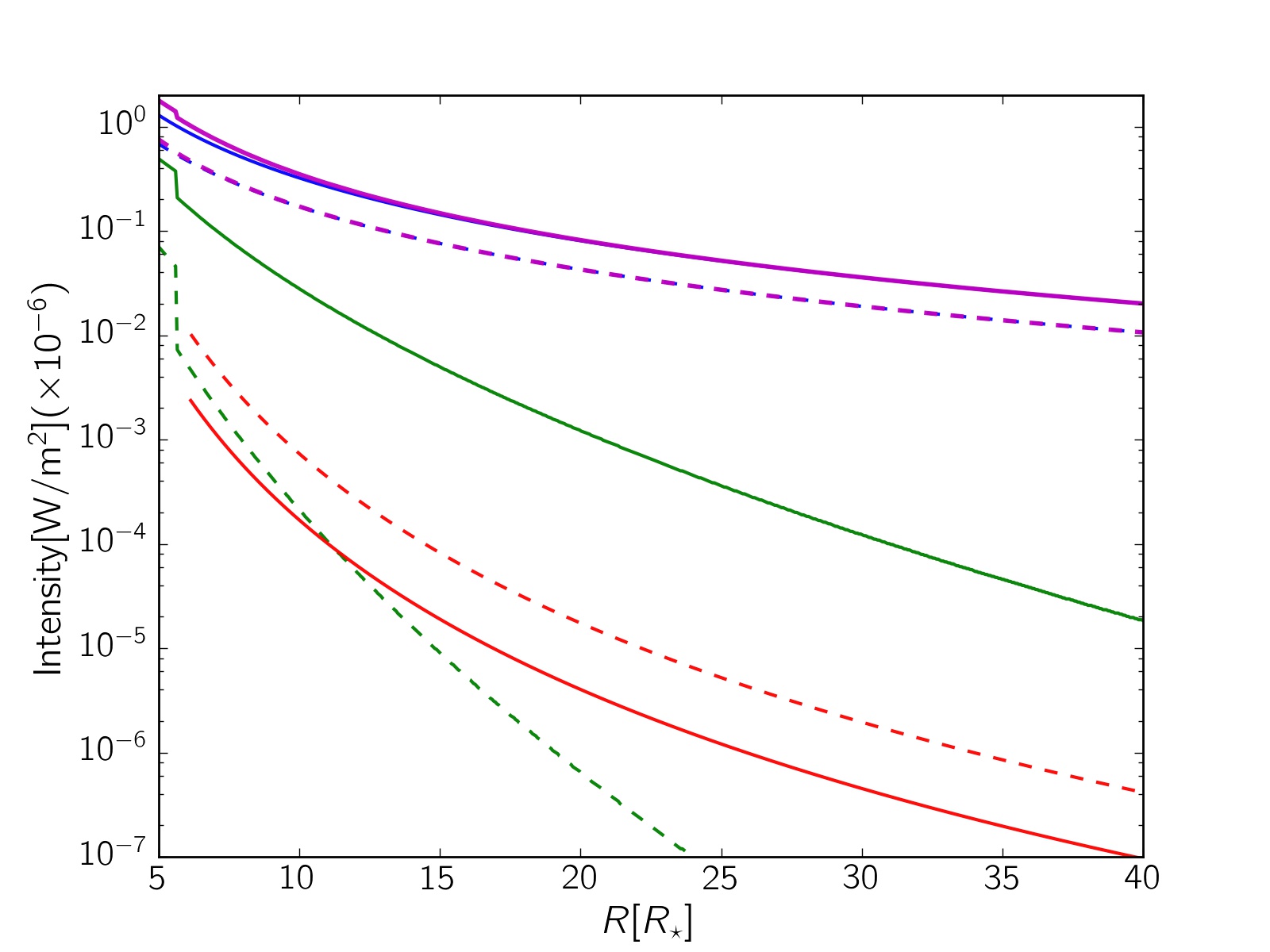}
    \caption{Left panel: The curves of luminosities of a Sun-like star (blue curve) and its disc which contains two components: thermal (green) and reflected (red) ones, versus wavelength. The overall luminosity due to the star and its disc is shown by the solid magenta curve. Right panel: Their intensity curves versus the radial distance in the midplane of the disc in the W149 (solid curves) and Z087 (dashed curves) filters. The parameters used for these plots are $M_{\textrm{disc}}= 1.4 \times 10^{-4}M_{\star}$,  $p=1.4$, $\beta=1.5$, $R_{\rm i}=4.7~R_{\star}$,  $R_{\rm f}=R_{\rm c}= 5.6~R_{\star}$.}
    \label{inten}
\end{figure*}

\section{Modeling disc}\label{two}

Our formalism to model the reflected and thermal emissions of discs around source stars is explained in this section. We assume that the discs studied here have cylindrical symmetry and are homogeneous without any planetesimal objects, magnetic fields, etc. A protoplanetary disc is generally divided into two regions: inner and outer regions. The inner region, extending from the initial radius $R_{i}$ to the outer radius $R_{f}$, contains pure and warm gas \citep{dullemond2010}. This part of the disc is optically thin. The outer region starts from the dust condensation radius $R_{\rm c}$, where the dust grains form and the effective temperature is $\sim1500~$K. This region continues to an outer radius $R_{\rm o}$. We set the outer radius where the disc surface density in the midplane reduces to $0.01$ of its maximum amount. The effective temperatures in inner and outer regions, $T_{i}$ and $T_{o}$, are related to the radial distance, 
\begin{eqnarray}
T_{i}&=& T_{\star} \sqrt{\frac{R_{\star}}{R}},\nonumber\\
T_{o}&=& T_{\star} \sqrt{\frac{R_{\star}}{R}} \epsilon^{-1/4},
\end{eqnarray}

\noindent where $T_{\star}$ and $R_{\star}$ are the effective temperature and the radius of the host source star, respectively. For the outer part, we take into account the warming of the dust grains by thermal radiation from the inner disc and wavelength dependence of dust opacity by considering the parameter $\epsilon$. This parameter is less than one for small dust grains \citep{dullemond2010}. The effective temperature of the inner region is in the range of $T_{i}(K) \in [2000, 2500]$, and this region corresponds to the distance range of $\sim0.02-0.04~$au for a Sun-like star. Between these two regions, an empty space can exist which is extended with the age of the disc.

The reflected emission from the outer disc is evaluated by solving the radial transfer equation \citep{chiang1997}. We determine the reflected fraction of the stellar radiation by the outer disc as \citep{simmons2002,Ignace2006},
\begin{eqnarray}
f_{\rm r}(R)=\frac{6 \sigma_{\rm{sc}}}{8} \int_{0}^{\infty} n(R, z)[(3 J-K) +(3K-J)(\frac{z}{r})^{2}] dz,
 \end{eqnarray}

\noindent where $n(R, z)= \rho(R, z)/m_{0}$ in the discs number density. $m_{0}$ is the mean molecular weight, $\rho$ represents the mass density, and $J$ and $K$ represent the first and second intensity moments, respectively. $\sigma_{\rm{sc}}$ is the Rayleigh scattering cross-section which is related to the light wavelength by $\sigma_{\rm{sc}}\propto \lambda^{-4}$. Considering the azimuthal symmetry, a disc density can be characterized in the cylindrical coordinates with two parameters the radial distance, $R$, in the equatorial plane and the vertical distance, $z$, from that plane. With this scheme, the mass density in the outer disc is
\begin{eqnarray}\label{eq11}
\rho(R,z)= \frac{\Sigma(R)}{\sqrt{2 \pi} H} \exp(-\frac{z^{2}}{2H^{2}}),
\end{eqnarray}

\noindent where $H= h_{0}(\frac{R}{R_{s}})^{\beta}$ is a growing function with the radial distance. $h_{0}$ and $R_{s}$ are the height and radial scales, respectively. $\Sigma(R)$ is the surface number density of dust grains and modeled as
\begin{eqnarray}\label{eq12}
\Sigma(R)=  \Sigma_{c} \left[\frac{R}{R_{c}}\right]^{-p},
\end{eqnarray}

\noindent here $\Sigma_{c}$ can be indicated according to the total mass of disc, $M_{\textrm{disc}}$, as
\begin{eqnarray}
M_{\textrm{disc}}= 2 \pi \int_{R_{c}}^{R_{o}} R dR \Sigma(R).
\end{eqnarray}

\noindent We note that $f_{r}$ is proportional to the scattering optical depth, i.e., $\tau_{sc}(R)= \Sigma(R) \sigma_{\rm{sc}}/m_{0}$. Finally, the reflected intensity is
\begin{equation}
I_{\rm{ref}}(R, \lambda)= f_{\rm r} I_{\star}(T_{\star}, \lambda) (\frac{R_{\star}}{R})^{2}.
\end{equation}

\noindent Here, $I_{\star}(T_{\star}, \lambda)$ is the intensity of the host star from its surface.
\begin{figure*}
    \centering
    \subfigure[]{\includegraphics[angle=0,width=0.49\textwidth,clip=0]{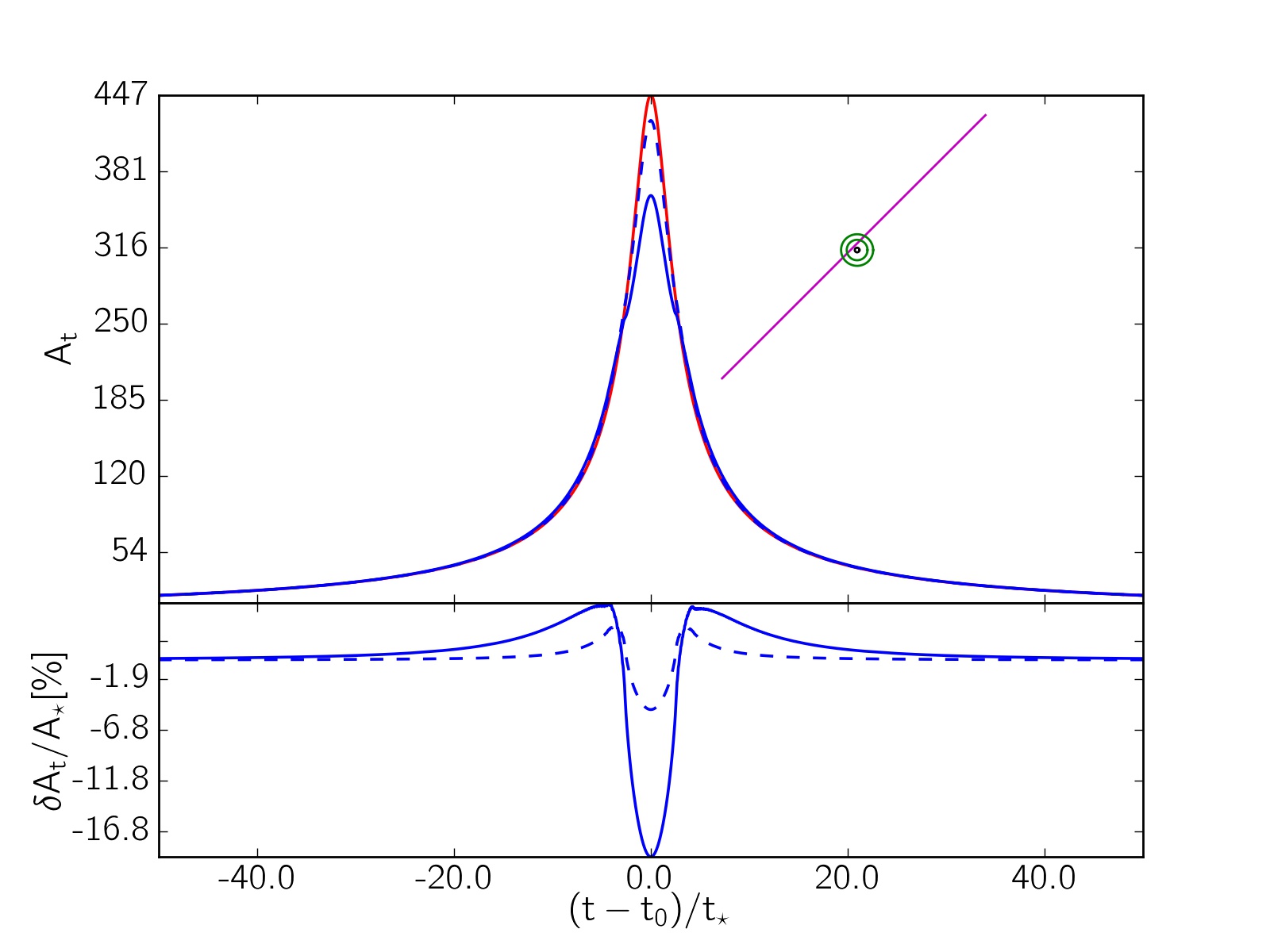}\label{fig2a}}
     \subfigure[]{\includegraphics[angle=0,width=0.49\textwidth,clip=0]{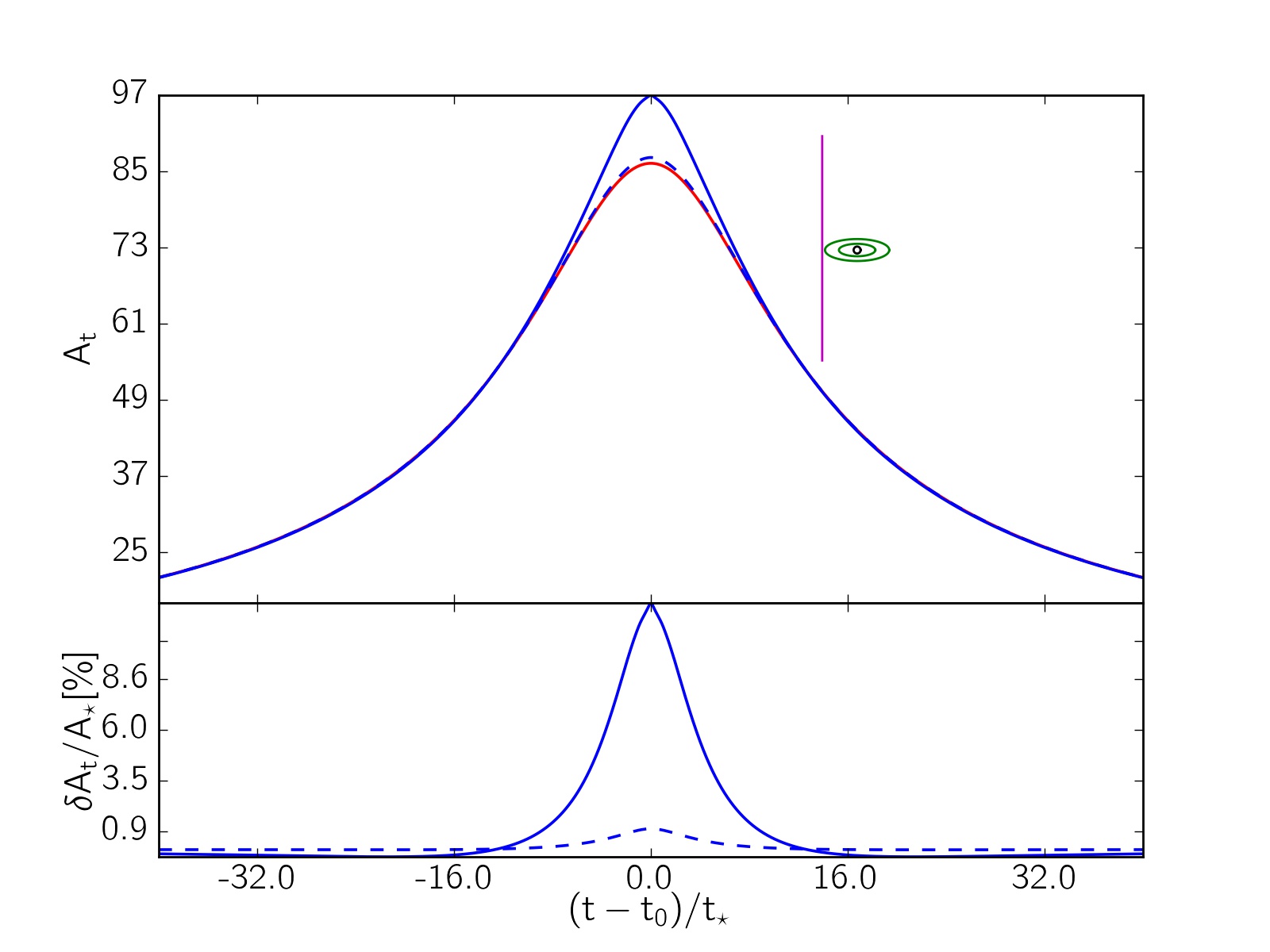}\label{fig2c}}
    \subfigure[]{\includegraphics[angle=0,width=0.49\textwidth,clip=0]{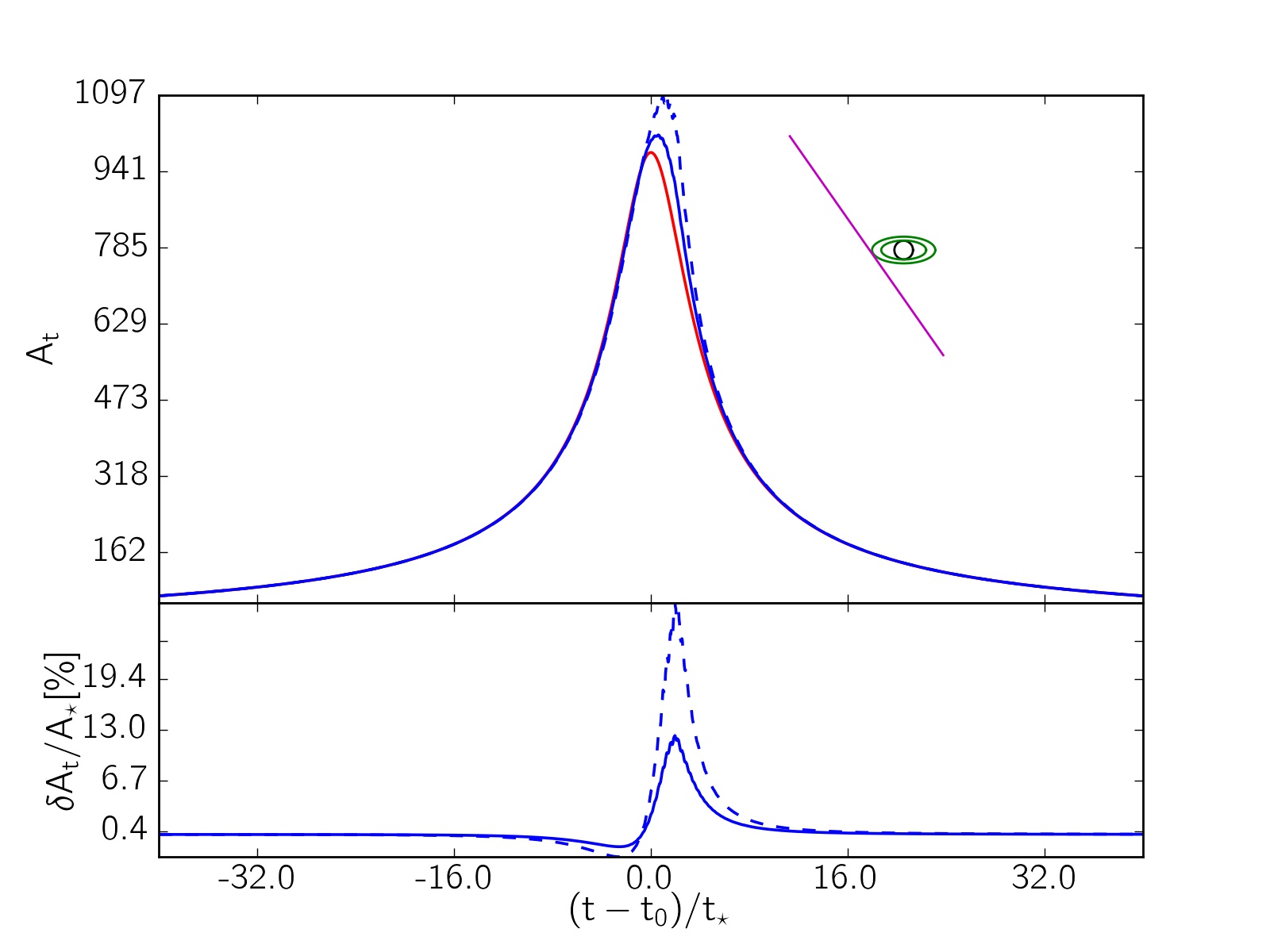}\label{fig2b}}
    \subfigure[]{\includegraphics[angle=0,width=0.49\textwidth,clip=0]{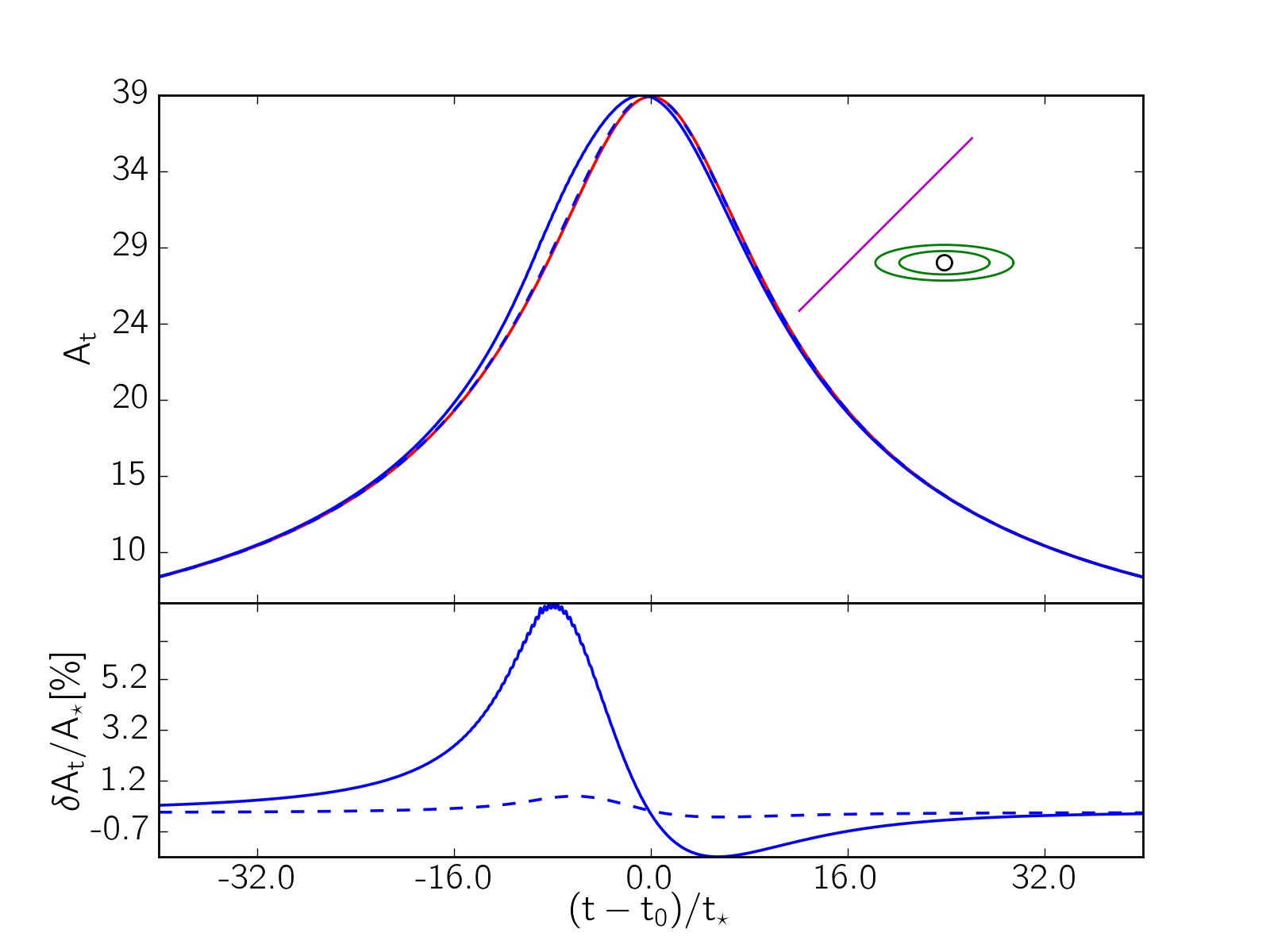}\label{fig2d}}
    \caption{Examples of microlensing light curves from source stars surrounded by discs. The solid and dashed blue curves are the light curves as seen in the W149 and Z087 filters, respectively. Their residuals in per cent with respect to the light curves of the source stars without discs (red curves) are plotted in the bottom panels. For each panel the inset displays the lens trajectories projected on the source plane (magenta straight lines), the source edge (black circle), and the inclined discs at the inner ($R_{i}$) and the condensation ($R_{\rm c}$) radii (green circles). The parameters used to make these light curves can be found in Table \ref{table}.} \label{light}
\end{figure*}

The intensity by the thermal emission from the disc is modeled by \citep{chiang1997,dullemond2001,Jami2015}:
\begin{eqnarray}
I_{\rm{ther}}(R,\lambda)= B(T(R),\lambda)[1-\exp(-\tau(R)/ \cos i)],
\end{eqnarray}

\noindent where $i$ is the inclination angle of the disc plane with respect to the sky plane. For a face-on disc, the inclination angle is zero. $B(T(R),\lambda)$ is the Planck distribution function at the temperature $T$ and the equatorial radial distance $R$. $\tau(R)=\Sigma(R)\kappa_{\lambda}$ is the optical depth at the radial distance $R$, $\kappa_{\lambda}$ is the dust opacity which is the absorption cross section of a photon with the wavelength $\lambda$ in the unit of the absorber mass. We calculate $I_{\star}$ according to the Planck distribution at the stellar temperature, i.e.,$$I_{\star}(T_{\star},\lambda)=B(T_{\star},\lambda)=\frac{8\pi h c^{2}}{\lambda^{5}}[\exp (\frac{hc}{k_{\rm B}T_{\star}\lambda})-1]^{-1}.$$\\

The energy densities of the disc and its host star are shown in Figure \ref{inten}. The left panel shows the relation between the luminosity and wavelength. The extra NIR bump due to the inner disc happens around $1-6~\mu$m. The reflected light has a little contribution in comparison to the thermal emission in near-infrared. The right panel shows the intensities in the unit of $\rm{W.m^{-2}}$ versus the radial distance in the disc midplane in the W149 (solid curves) and Z087 (dashed curves) filters, which will be utilized by \wfirst.~The parameters used for these two plots are $\log_{10}[M_{\textrm{disc}}/M_{\star}]=-3.85$, $p=1.4$, $\beta=1.5$, $R_{\rm i}=4.7~R_{\star}$,  $R_{\rm f}=R_{\rm c}= 5.6~R_{\star}$. Accordingly, the maximum disc emission in NIR is due to the thermal component radiated from the inner disc. This radiation is more dominant in W149 than Z087. The reflected radiation from the disc in the Z087 filter is more than that in W149 because its mean wavelength, i.e., $0.87~\mu m$, is shorter than the mean wavelength of W149. In addition, the cross-section of the Rayleigh scattering is larger in the Z087 filter than that in the W149 filter.

\section{Microlensing of source stars surrounded by discs}\label{three}

The detection of distant discs through microlensing was investigated in several references. Briefly, the disc detection in microlensing observations is possible through three channels. (i) Detecting the gravitational perturbation effects of circumstellar discs around microlenses \citep{Bozza2002a, Bozza2002b, Hundertmark2009}. (ii) Detecting the mid- and far-infrared extra emission at the wavelength $\sim20~\mu$m from discs surrounding the source stars in single microlensing events \citep{zheng2005}. (iii) Detecting inclined circumstellar discs around source stars through polarimetry observations of high-magnification microlensing events \citep{sajadian2015}. The reflected light of the source star from an inclined disc has a net polarization signal which depends on the disc inclination angle.

\begin{figure*}
	\centering
	\subfigure[]{\includegraphics[angle=0,width=0.49\textwidth,clip=0]{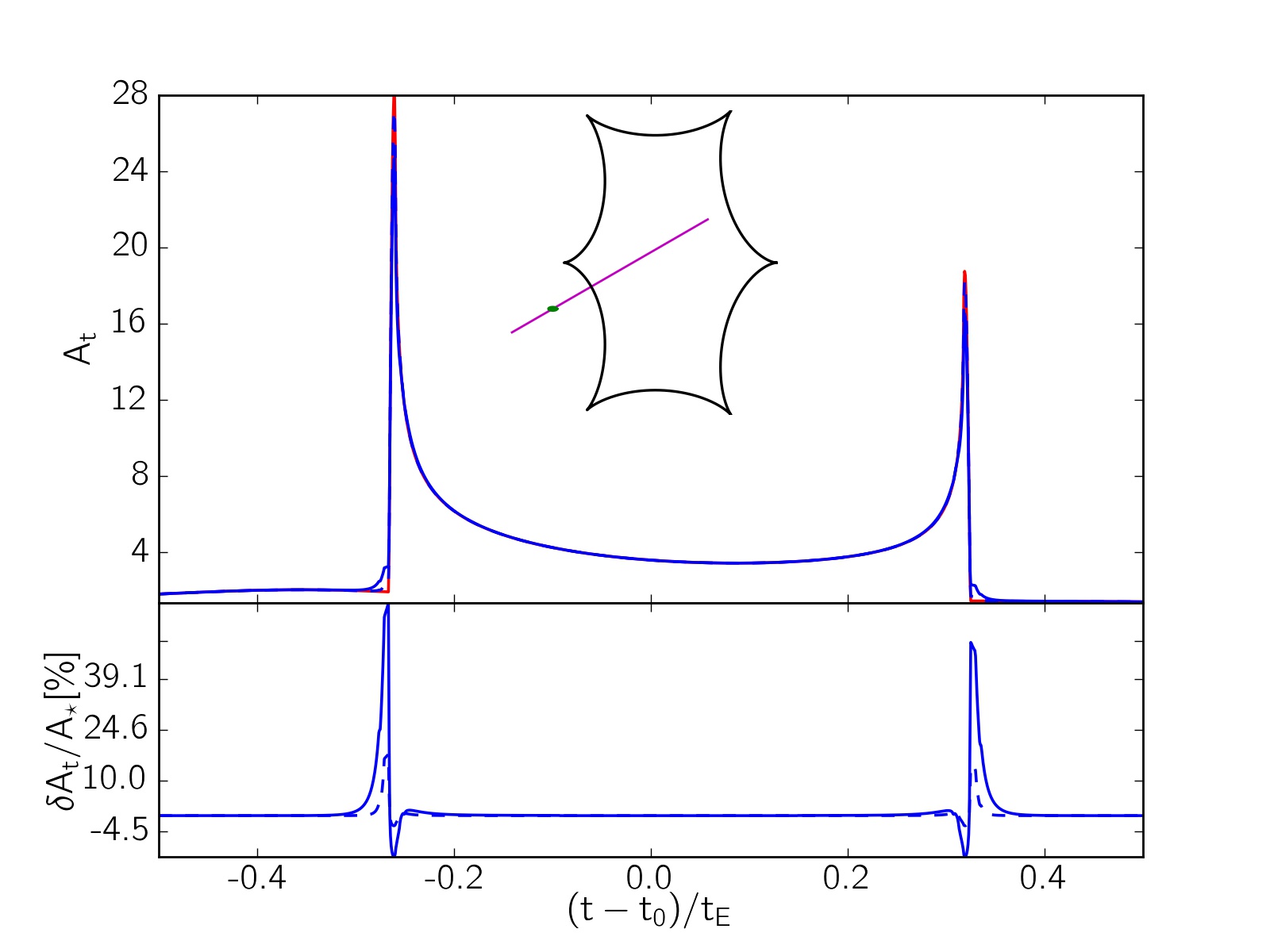}\label{fig3a}}
	\subfigure[]{\includegraphics[angle=0,width=0.49\textwidth,clip=0]{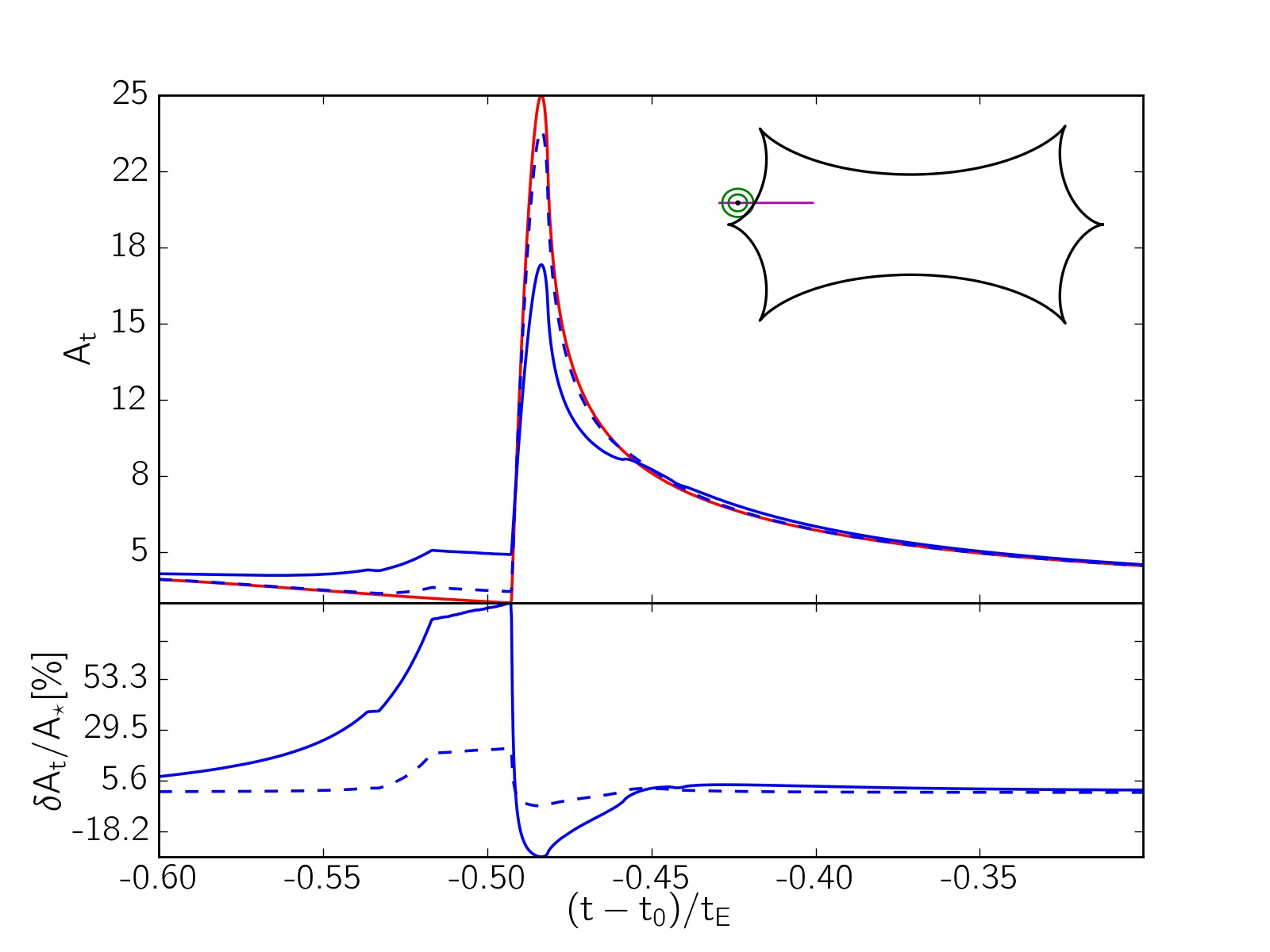}\label{fig3b}}
	\subfigure[]{\includegraphics[angle=0,width=0.49\textwidth,clip=0]{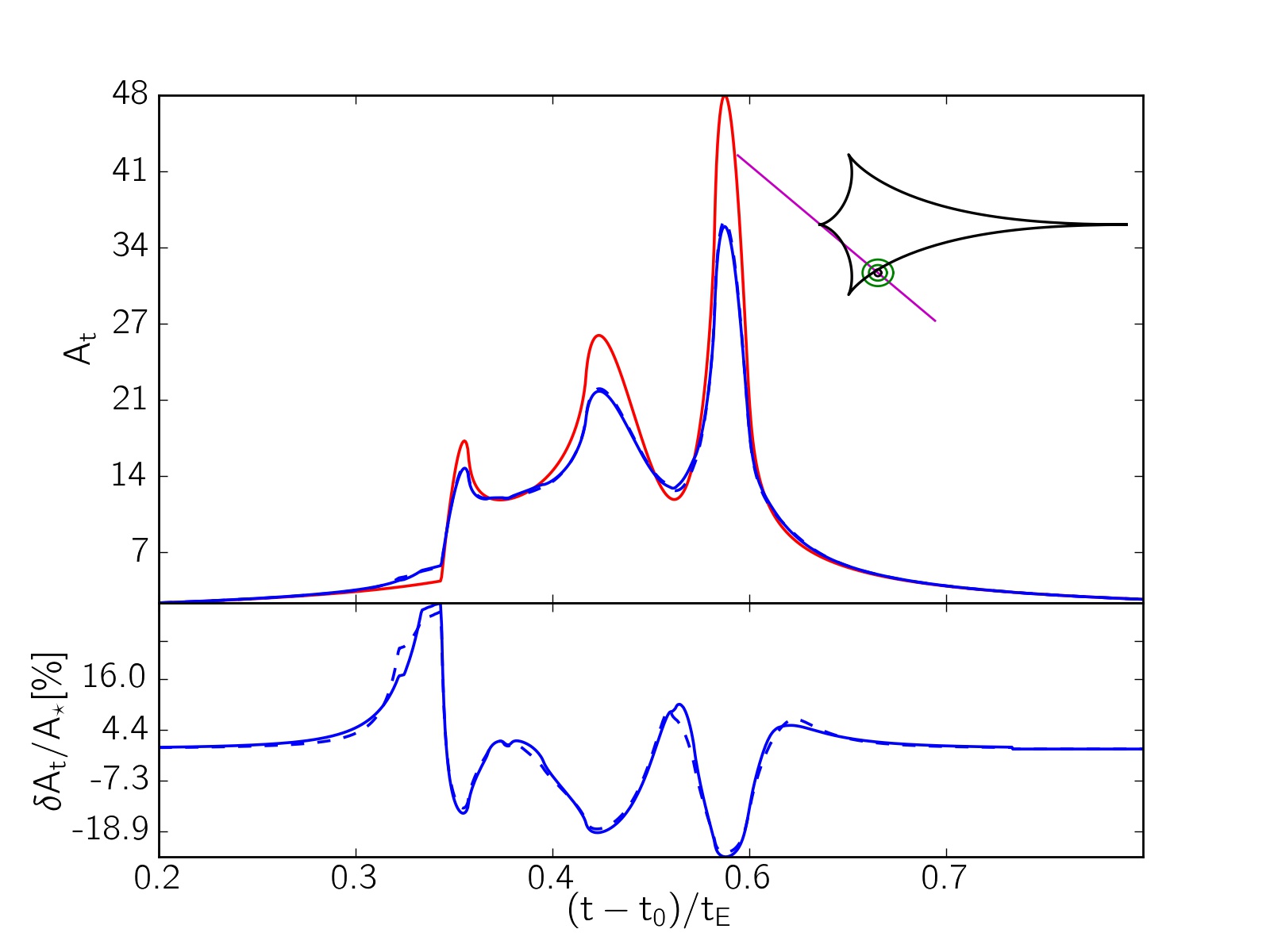}\label{fig3c}}
	\subfigure[]{\includegraphics[angle=0,width=0.49\textwidth,clip=0]{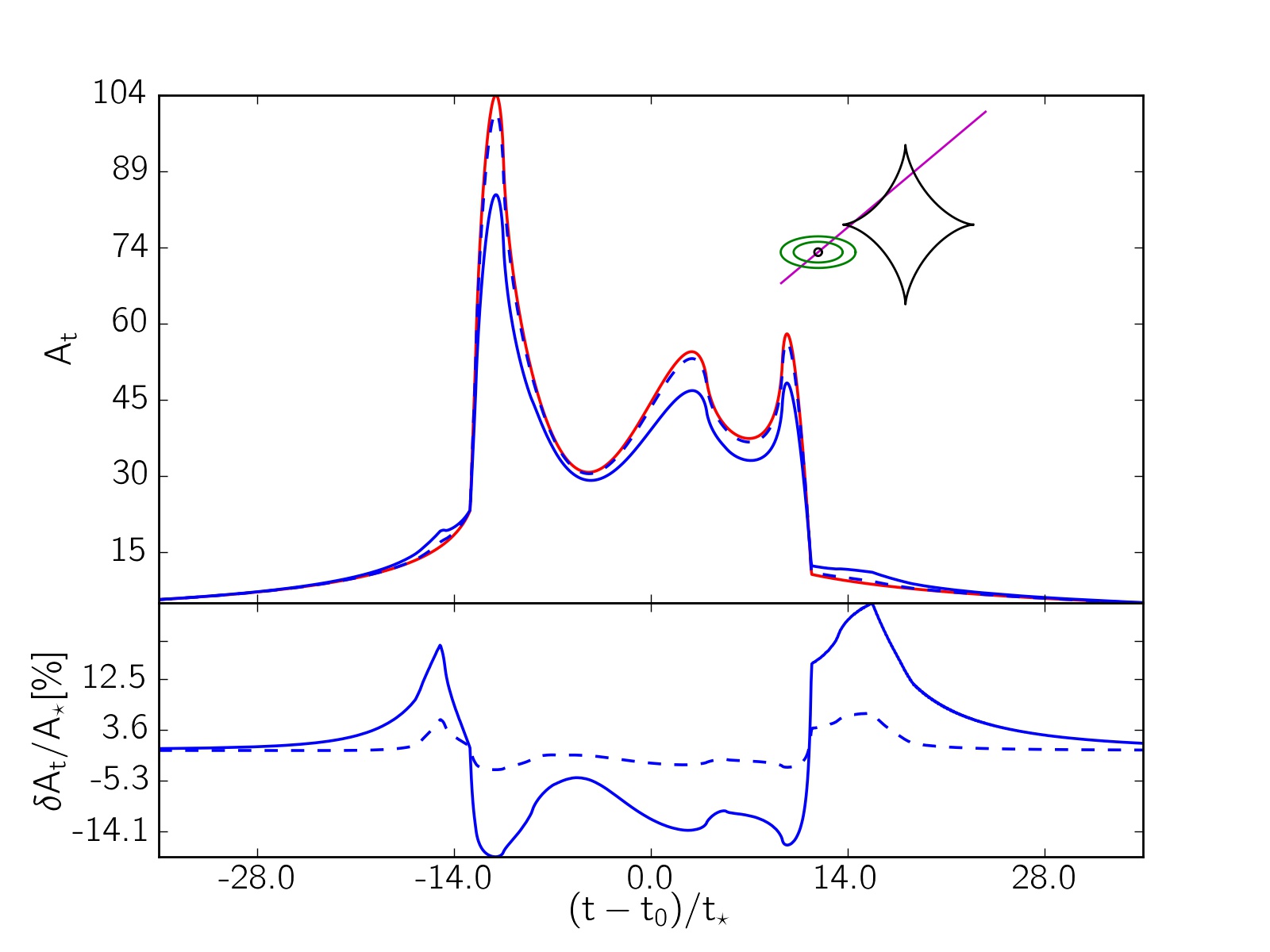}\label{fig3d}}
	\caption{Examples of caustic-crossing binary light curves from source stars surrounded by discs. The caustic curves projected on the source plane are represented by solid black curves inside the inset parts. More details about the plots can be found in the caption of Figure \ref{light}.}\label{fig3}
\end{figure*}

In this section, we investigate the disc-induced perturbations in single and binary microlensing light curves in both filters of \wfirst.~The magnified and normalized luminosity received by the observer from the disc is determined by integrating its intensity over the disc surface and the wavelength, as

\begin{eqnarray}\label{ldisk}
L_{\textrm{disc}, \rm F}= \int_{0}^{\infty}d\lambda~T_{\rm F}(\lambda) \int dx\int dy~[I_{\rm{ref}} + I_{\rm{thre}}]A_{\rm d}(x,y),
\end{eqnarray}

\noindent where $T_{\rm{F}}(\lambda)$ is the throughput function for the passband filter $\rm{F}$ ($\rm{F}$ is either W149 or Z087),  $A_{\rm{d}}(x,~y)$ is the magnification factor of each element of disc at the projected position of $(x,~y)=(R \cos \theta,~R \sin \theta~\cos i)$, where $\theta$ is the azimuthal angle over the disc. The magnification factor, $A_{\rm{d}}$, depends on the lens position projected on the source plane. The total magnification factor due to the source star and its disc is

\begin{eqnarray}\label{atotal}
A_{\rm t}=\frac{L_{\star, \rm F}~A_{\star}(u, \rho_{\star})+L_{\rm{disc}, \rm F}}{L_{\star, \rm F} + L_{\rm{disc}, \rm F, 0}},
\end{eqnarray}

\noindent where $u$  is the lens-source distance projected on the lens plane and normalized to the Einstein radius, i.e., the radius of the image ring when the lens, source, and the observer are completely collinear. $\rho_{\star}$ is the source radius projected on the lens plane and normalized to the Einstein radius, and $L_{\rm{disc}, \rm F, 0}$ is the intrinsic disc luminosity without lensing effects. The intrinsic stellar luminosity, $L_{\star,~\rm F}$, is
\begin{eqnarray}
L_{\star, \rm F}(T_{\star})=\pi R_{\star}^{2} \int_{0}^{\infty}d\lambda~T_{\rm F}(\lambda)~I_{\star}(T_{\star},\lambda).
\end{eqnarray}

\noindent Single and binary microlensing events of source stars with discs are studied in the following subsections.

\begin{figure*}
\centering
\subfigure[]{\includegraphics[angle=0,width=0.49\textwidth,clip=0]{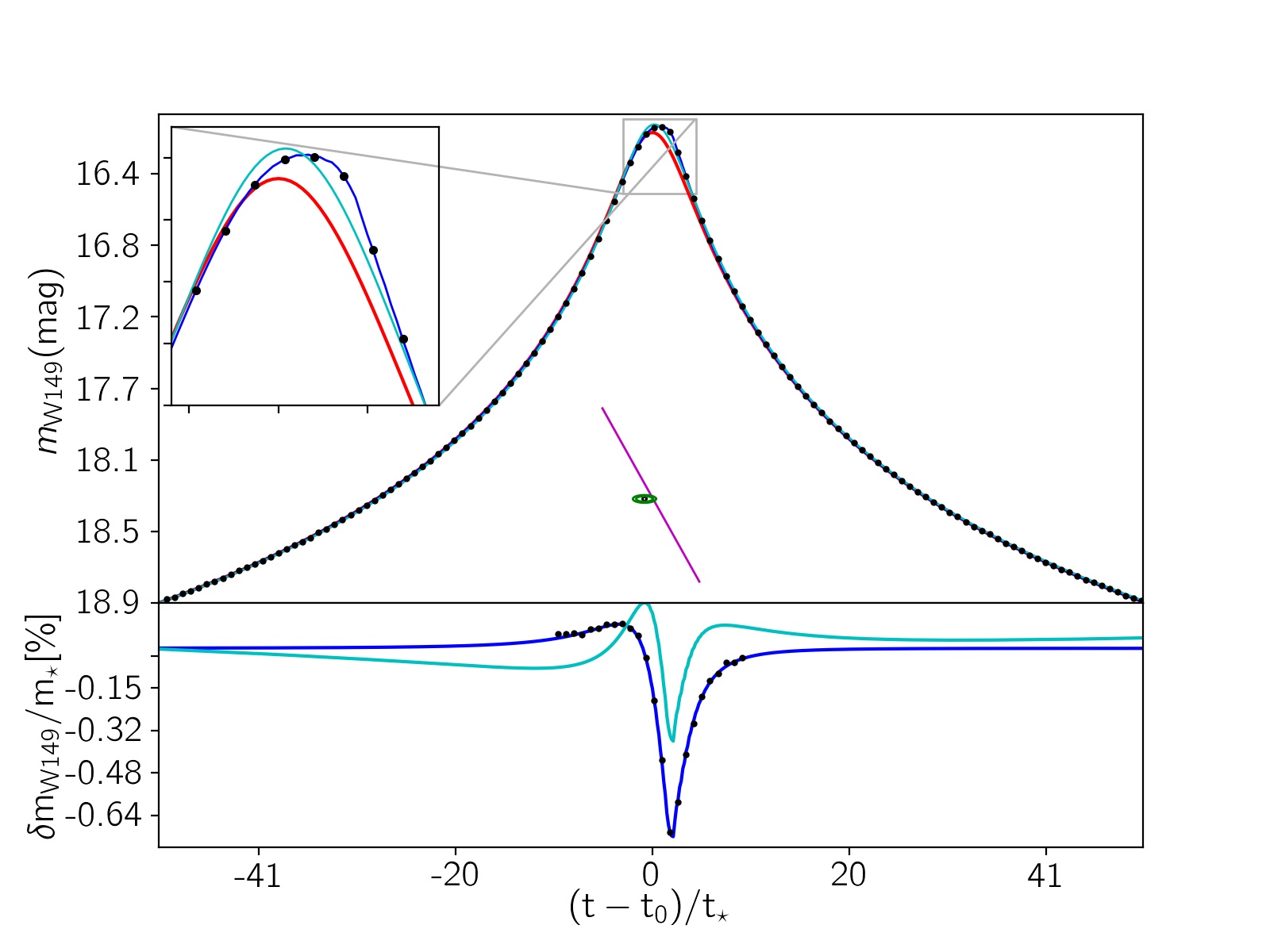}\label{fig4a}}
\subfigure[]{\includegraphics[angle=0,width=0.49\textwidth,clip=0]{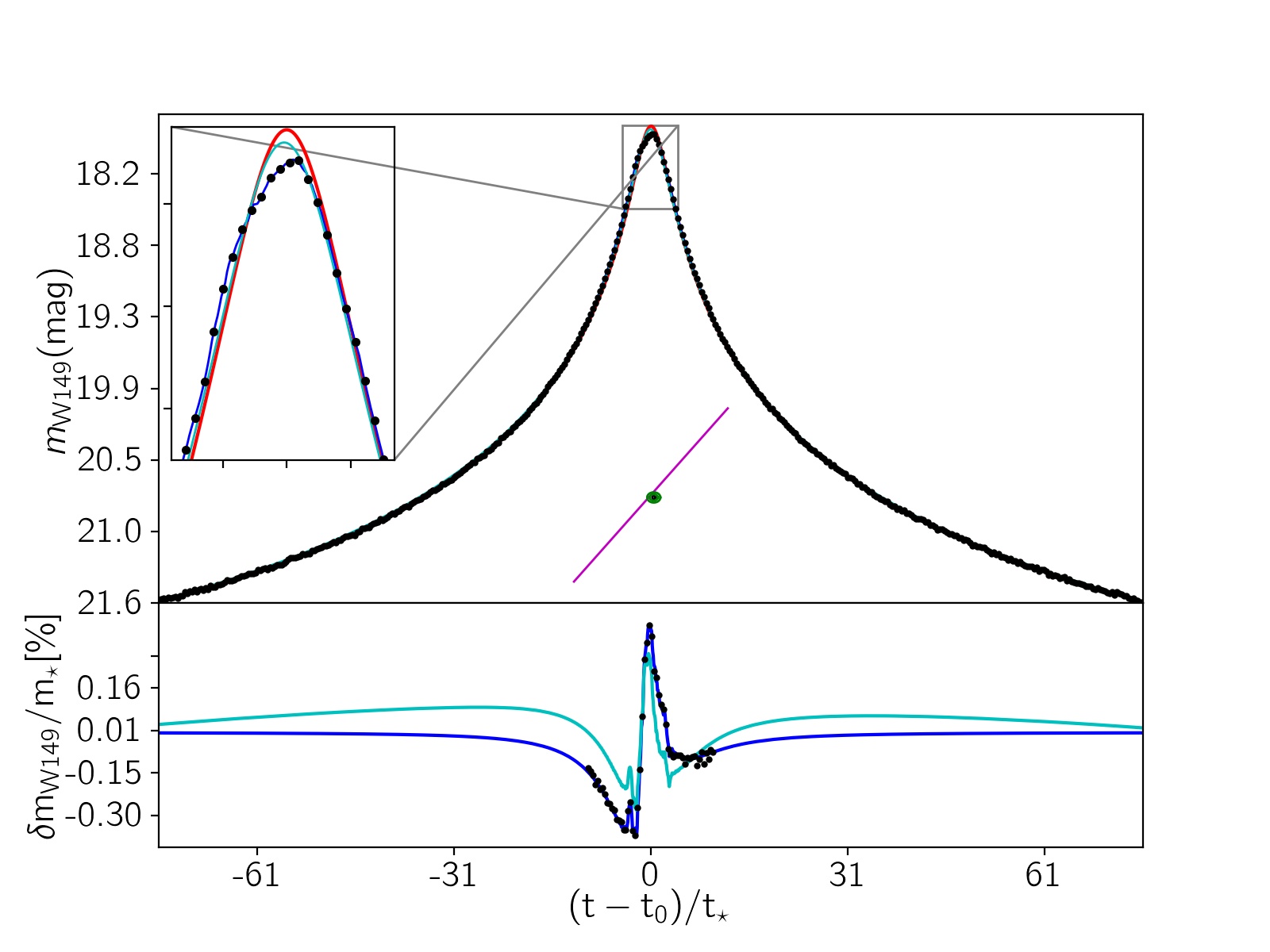}\label{fig4b}}
\subfigure[]{\includegraphics[angle=0,width=0.49\textwidth,clip=0]{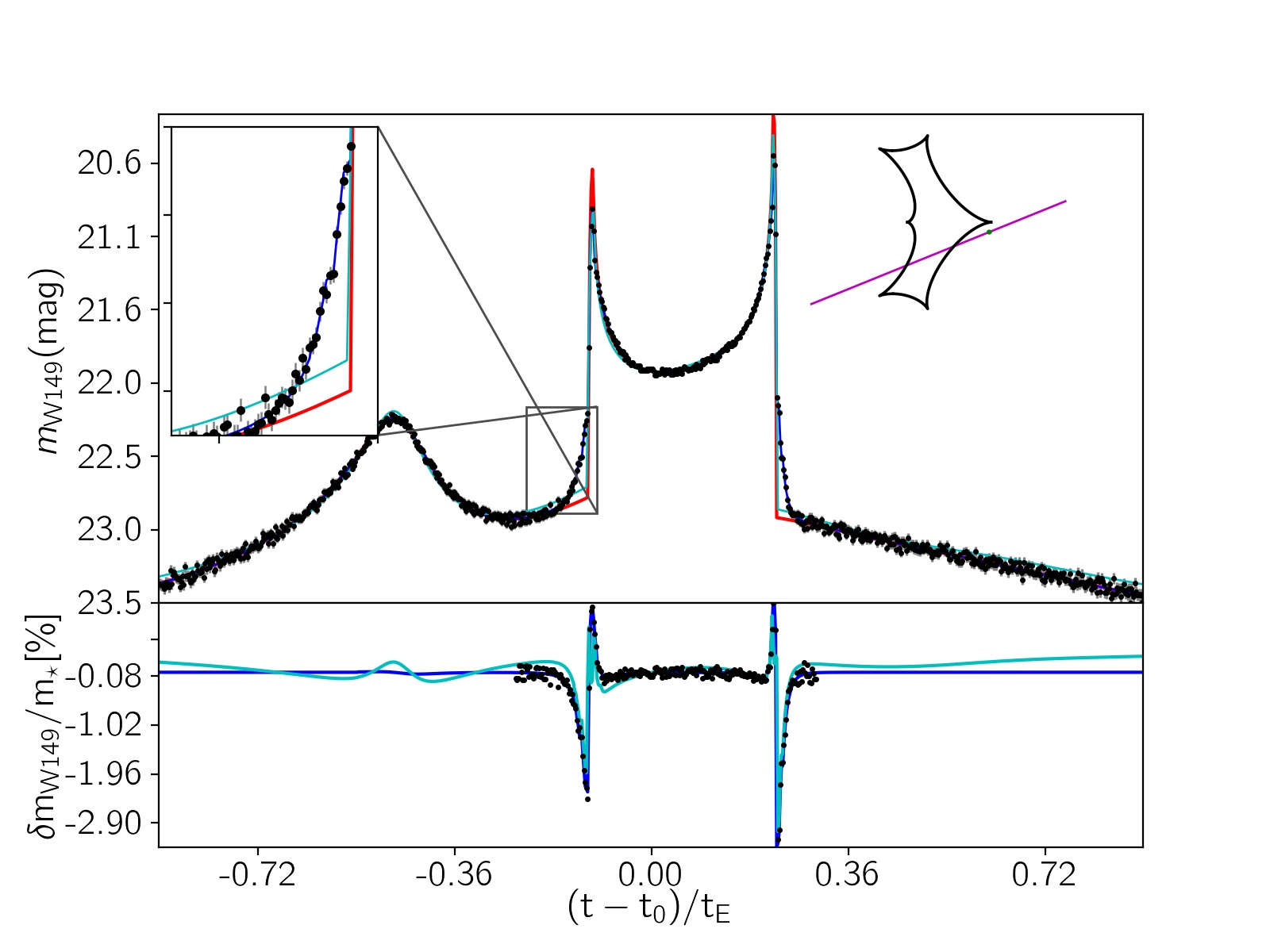}\label{fig4c}}
\subfigure[]{\includegraphics[angle=0,width=0.49\textwidth,clip=0]{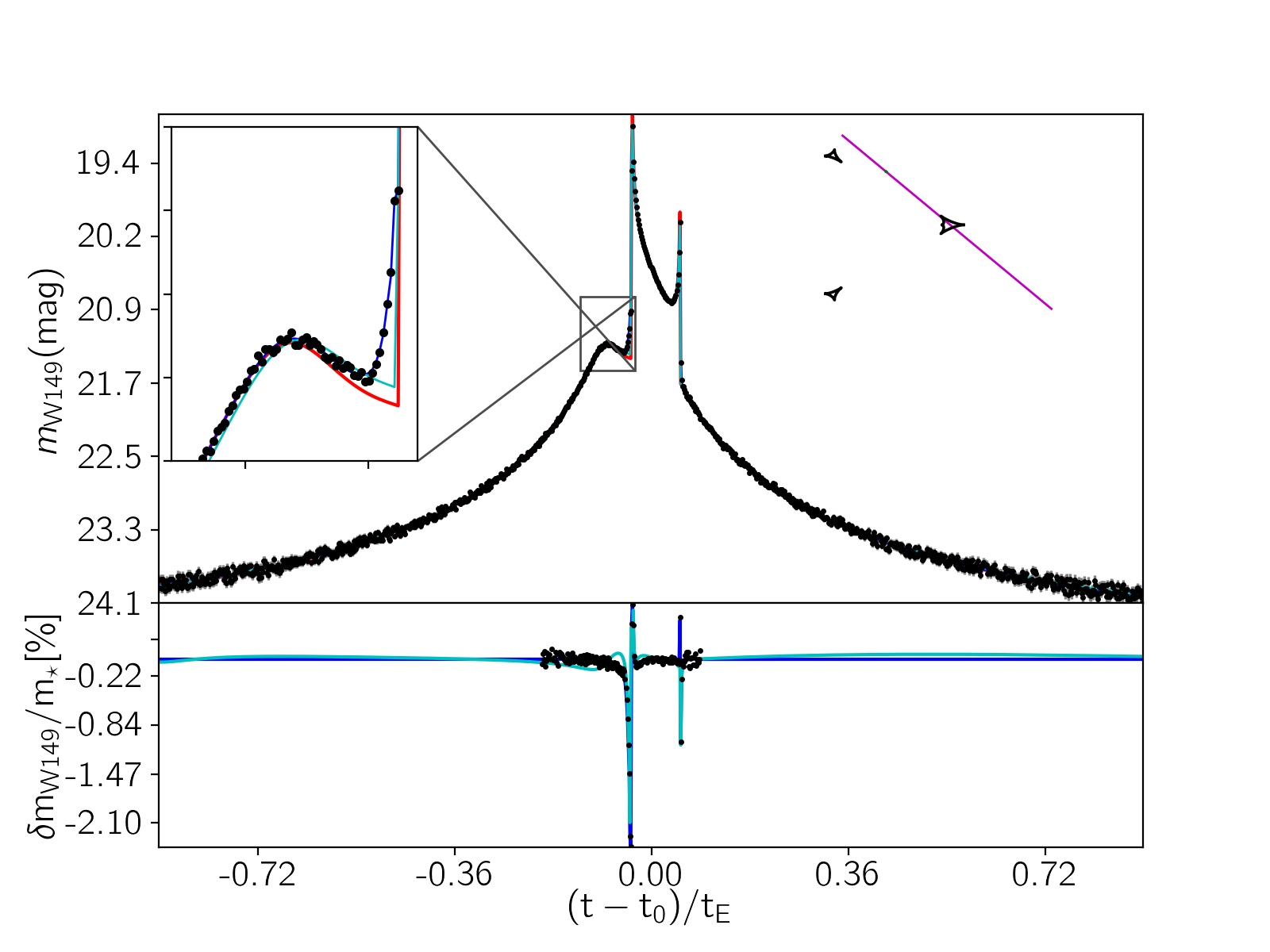}\label{fig4d}}
\caption{Four examples of simulated microlensing light curves of the source stars with circumstellar discs. In these plots, synthetic data points as seen by \wfirst~in the W149 filter are shown by black filled circles. The red and blue solid curves show the magnified magnitude of the source star itself (Equation \ref{mstar})  and the source star surrounded by the disc (Equation \ref{mw149}) versus time, respectively. The best-fitted light curves to the data points are represented with cyan curves. The parameters used to make them and $\Delta \chi^{2}_{\rm{d}}$ values are reported in Table \ref{table}. In all of these events, the disc perturbations are discernible.}\label{fig4}
\end{figure*}

\subsection{Single microlensing}

The effect of the disc on the microlensing light curves in near-infrared depends strongly on (i) the lens impact parameter, $u_{0}$, compared with the source and disc radii projected on the lens plane and normalized to the Einstein radius, i.e., $\rho_{\star}$, $\rho_{\rm i}$, $\rho_{\rm c}$, and (ii) the angle between the lens trajectory with respect to the semi-major axis of the disc (symmetry or asymmetry). In Figure~\ref{light} four examples of microlensing light curves due to the source stars surrounded by discs are shown. In each plot, the magnification factor without disc effects is shown by the red solid curve. The total magnification factors due to the disc and its host star in the W149 and Z087 filters are represented by blue solid and dashed curves, respectively. The residuals in the total magnification factor, $\delta A_{\rm{t}}/A_{\star}= (A_{\rm{t}}-A_{\star})/A_{\star}$, in per cent in these filters are plotted in the bottom panels. The parameters of each light curve are reported in Table~\ref{table}. In order to classify the perturbations, we separate two geometric configurations, symmetric or asymmetric, in the following. \\

\textbf{(i):~}If the lens trajectory is parallel with (or normal to) the disc semi-major axis or if the disc is almost face on, the light curve will be symmetric with respect to the time of closest approach. So both sides of the light curve will be similar. In that case, the disc-induced perturbations for different values of the lens impact parameter are classified as \\

\noindent(a)~If $u_{0}<\rho_{\rm{i}}$, two extra (symmetric and very flattened) peaks are produced while the lens is transiting the discs inner radius and the maximum magnification at the time of closest approach is flattened (demagnification happens). A sample light curve is shown in Figure \ref{fig2a}.\\

\noindent(b)~If $\rho_{\rm{i}}<u_{0}<\rho_{\rm{c}}$, the maximum magnification at the time of closest approach increases. Because two peaks due to the source star and inner disc coincide with each other. In that case, discerning the disc from the resulted light curve is impossible. \\

\noindent(c)~If $u_{0}>\rho_{\rm{c}}$, the disc-induced perturbation on the light curve is very small and it slightly enhances its maximum value. An example light curve for this situation is represented in Figure \ref{fig2c}. Comparing the scales on $y-$axes in the residual panels of Figures \ref{light} reveals that the disc perturbations reduce by enlarging the lens impact parameter. \\

\textbf{(ii):~}If the lens trajectory is not parallel with or normal to the semi-major axis of the disc and its inclination angle is not zero, the resulted light curve is not symmetric with respect to the time of closest approach. This symmetry breaking is very helpful for discerning discs around source stars. In that case, there are three ranges for the lens impact parameter. \\

\noindent(a)~If $u_{0}<\rho_{\rm{i}}$, two asymmetric disc-induced peaks form in both sides of the main peak. When the lens reaches the empty region between the source star and the disc inner edge a small demagnification occurs.\\

\noindent(b)~If $\rho_{\rm{i}}<u_{0}<\rho_{\rm{c}}$, the peak of the light curve is deformed because an extra peak due to disc is created very close to the main peak of the light curve, see, e.g., Figure \ref{fig2b}. Owing to very small distance between these peaks, in the resulted light curve the main peak is not at the time of closest approach. In this plot, the disc perturbation in Z087 is larger than that in W149. When the lens is passing close to the disc condensation radius (where the reflected radiation from the disc maximizes) the perturbation in the Z087 filter is larger than that in the W149 filter.\\

\noindent(c)~If $u_{0}>\rho_{\rm{c}}$, a very extended peak happens at the time of closest approach of the lens and disc inner radius, see, Figure \ref{fig2b}. This time is not $t_{0}$. Because the lens passes close to an ellipse instead of a circle. Comparing figures \ref{fig2c} and \ref{fig2d} manifests the displacement in the peak position in asymmetric configuration of disc and lens trajectory.\\

\noindent Totally, the discernible effect of a disc around the source star in the single microlensing light curve is the appearance of two extra, very extended and asymmetric peaks in both sides of the main peak when the lens transits the disc inner edge. The asymmetry in the microlensing light curves when the lens impact parameter is larger than the inner disc radius is the best signature for the existence of discs around the source star. The effect of a disc on the overall magnification of the disc and its host source star differs significantly in the W149 and Z087 filters. Because the disc emission maximizes in infrared and is proportional to the radial distance from the disc centre.\\

\subsection{Binary caustic-crossing microlensing}

Here, we study the deviation due to a disc around the source star in the binary lensing events with caustic-crossing features. Four examples of binary microlensing light curves of the source stars surrounded by discs are represented in Figure \ref{fig3}. The characterizations of these light curves are similar to ones plotted in Figure \ref{light}. In the insets, caustic curves are added and shown by black solid curves. The parameters of these light curves can be found in Table \ref{table}. \\

\noindent According to these light curves, the largest deviations due to discs on the binary light curves happen right before entering or immediately after exiting from the caustic curve by making extra peaks. In addition, the disc decreases the maximum magnification while crossing the caustic by increasing the finite size effect, see, e.g., Figure  \ref{fig3a}. If the caustic size is on the order of the disc size, the disc perturbation decreases significantly. This result is inferred from comparing the y-scales in the residuals panels of Figures \ref{fig3a} and \ref{fig3d}. According to Figures \ref{fig3b} and \ref{fig3c}, while caustic crossing a disc with larger inner radius generates bigger peaks located at further distances from the main peak. Hence, discs with larger inner radii have more chance to be discerned. The best criterion for the existence of discs is the extra peaks close to the main peaks. We note that an extra solar planet around the source star can not make such pair peaks in both locations. The perturbations due to exoplanets are not generally symmetric. The disc perturbation when the source is inside the caustic curve is too small to be detected. Hence, the disc signatures in caustic-crossing microlensing events can be probed right before and immediately after caustic-crossing futures.
\begin{table*}
\centering
\caption{The parameters for generating sample microlensing light curves shown in Figures \ref{light}, \ref{fig3} and \ref{fig4}. The parameter $\alpha$ is the angle between the source trajectory with respect to the horizontal axis (which is the binary axis in binary microlensing). $m_{\rm b}$ is the baseline magnitude related to the overall flux due to the source and blending stars in the W149 filter.}\label{table}
\begin{tabular}{cccccccccccc} 
\hline
& $i$ & $\log_{10}[M_{\rm{disc}}/M_{\star}]$  & $R_{i}$ & $R_{c}$ & $\log_{10}[\tau_{\rm{sc}}(R_{c})]$ &  $u_{0}$ & $\rho_{\star}$ & $t_{\rm{E}}$ & $\alpha$ &  $m_{\rm b}$ & $\Delta \chi_{\rm d}^{2}$ \\
& $(\rm{deg})$ & $~$ & $(R_{\star})$  & $(R_{\star})$ & $~$ & $~$ & $~$ & $\rm{(day)}$ & $\rm{(deg)}$ & $\rm{(mag)}$ &  \\
\hline
\ref{fig2a} & $10.0$  & $-4.5$ &  $2.96$ &   $4.58$ &    $-0.37$ &  $0.002$ &  $0.001$ & $10.0$ & $45.0$, &  $22.1$  &  $-$\\
\ref{fig2b} &  $65.0$ &  $-4.0$ & $2.23$ &   $3.13$ &    $0.89$  &  $0.001$  & $0.0003$&$21.0$ & $125.0$ &  $24.04$ & $-$\\
\ref{fig2c}  & $70.0$ &  $-5.0$ &  $4.37$ &   $7.72$ &    $-1.82$ & $0.012$ &  $0.001$ & $13.0$& $90.0$ &    $20.52$ & $-$ \\
\ref{fig2d} &  $75.0$ &  $-5.0$ &  $5.00$ &   $7.65$  &    $-1.62$ &  $0.026$  & $0.003$&$15.0$& $45.0$ & $19.73$ & $-$ \\\
\ref{fig3a} &  $65.0$ &  $-5.0$ & $2.59$ &   $4.47$ &    $-0.31$  & $0.05$  &  $0.003$    & $10.0$& $30.0$ &  $22.48$ & $-$ \\
\ref{fig3b} &  $25.0$  & $-4.3$ &  $5.26$  &  $8.82$ &   $-0.89$  & $0.07$  &  $0.005$    & $10.0$& $0.0$  & $19.85$ & $-$ \\
\ref{fig3c} & $30.0$ & $-4.0$ &  $2.62$ &  $4.47$  &   $0.98$ &  $0.46$     &  $0.008$    & $8.0$  & $140.0$ & $22.42$ &  $-$ \\
\ref{fig3d}&  $65.0$ &  $-4.0$ &  $5.33$ & $8.15$ &  $-0.41$ &  $0.04$ &  $0.004$ &  $8.0$  & $40.0$ & $19.57$ &  $-$  \\
\ref{fig4a} &  $70.5$ &  $-7.49$ &  $3.21$ &  $4.48$ &   $-3.31$ &  $0.008$  & $0.002$ &  $5.95$ & $299.2$ & $21.36$ &  $6540.8$ \\
\ref{fig4b}& $38.6$ &  $-6.58$ &  $2.59$ &  $4.25$ &   $-2.86$  & $0.008$ &  $0.003$ &   $8.84$ & $48.4$ &  $23.11$ & $6560.1$\\
\ref{fig4c}& $16.5$ &  $-5.95$ &  $2.27$ &  $4.42$ &   $-1.18$  & $0.216$ &  $0.002$ &   $4.51$ & $202.1$ &  $23.8$ & $11142.3$\\
\ref{fig4d}& $59.7$ &  $-6.03$ &  $2.14$ &  $3.69$ &   $-2.22$  & $0.02$ &  $0.0007$ &   $6.50$ & $320.3$ &  $24.4$ & $6980.4$\\
\hline
\end{tabular}
\end{table*}

\section{\wfirst~capability for disc detection in microlensing observations}\label{four}

In order to study the \wfirst~ability for discerning the disc-induced perturbations, we simulate the synthetic data points taken by this telescope in the W149 filter. Then, we investigate if these signatures are distinguishable from fitting two microlensing models, the real model with disc and a best-fitted model without discs. The cadence of data taken in W149 is fixed at $15.16~$min. We ignore the data will be taken in the Z087 filter. Because, their cadence, i.e., $12$h, is too long and as a result the number of data points in this filter is low. In addition, because of the different radiations of the disc in these filters, their light curves are not similar and can not be converted to each other only by shifting. We consider three 72-day seasons at the first of the \wfirst~ mission and three similar at the end of it \citep{spergel2015}. The error bar in the apparent magnitude of the source star as measured by \wfirst~in the W149 filter increases by enhancing the source magnitude \citep[according to Fig (4) of][]{Penny2019}. The error in the magnification factor as a function of the error in the magnitude is given by
\begin{eqnarray}
\delta_{\rm A}= 10^{-0.4 \delta_{\rm m}}-1,
\end{eqnarray}
where, $\delta_{\rm{m}}$ and $\delta_{\rm{A}}$ are the errors in the apparent magnitude and the magnification factor, respectively.\\

In order to study the statistic of the discs which can be detected in microlensing observations by \wfirst,~we generate a big ensemble of microlensing events towards the Galactic bulge by considering the relevant distribution functions of the lens and the source stars. Many details about this simulation can be found in the previous papers \citep{Sajadian2019, Moniez2017, bagheri2019}. We indicate the absolute magnitude of the source stars in W149 as $M_{\rm{W149}}= \rm{(H+J+K)}/3$, where H,~J,~K are their absolute magnitudes in the standard Johnson-Cousin photometry system \citep{Montet2017}. The magnified apparent magnitude of the source star and its disc is
\begin{eqnarray}\label{mw149}
m_{\rm{W149}}=  m_{\rm{b}} -2.5 \log_{10}[b~A_{\rm t} +1-b],
\end{eqnarray}
where $m_{\rm{b}}$ is the baseline apparent magnitude in W149 related to the overall light entering the \wfirst~PSF (from the source and its neighborhood blending stars), $b$ is the blending factor. $A_{\rm{t}}$ is the total magnification factor and given by Equation \ref{atotal}.\\

For simulating the binary systems as binary microlenses we choose the semi-major axes of the binary orbits from the \"Opik's law \citep{Opik}. The mass ratio of two components, $q$, is taken from the Gaussian distribution function with the mean value of $0.23$ which was given by \citet{1991Duq}.\\

For simulating discs, the disc mass is uniformly chosen in the logarithmic scale in the range of $\log_{10}(M_{\rm{disc}}/M_{\star})\in[-5.5,~-8.5]$. The powers $\beta$ and $p$ (in Equations \ref{eq11} and \ref{eq12}) are chosen uniformly from the range $[1.25,~1.55]$. We fix the edges of the inner disc, i.e., $R_{\rm i}$ and $R_{\rm f}$, where the temperature reaches to $2500$ and $2000~$K, respectively. The height scale $h_{0}$ is chosen uniformly from the range $[8,~12]$~au.\\
\begin{figure*}
\centering
\includegraphics[angle=0,width=0.49\textwidth,clip=0]{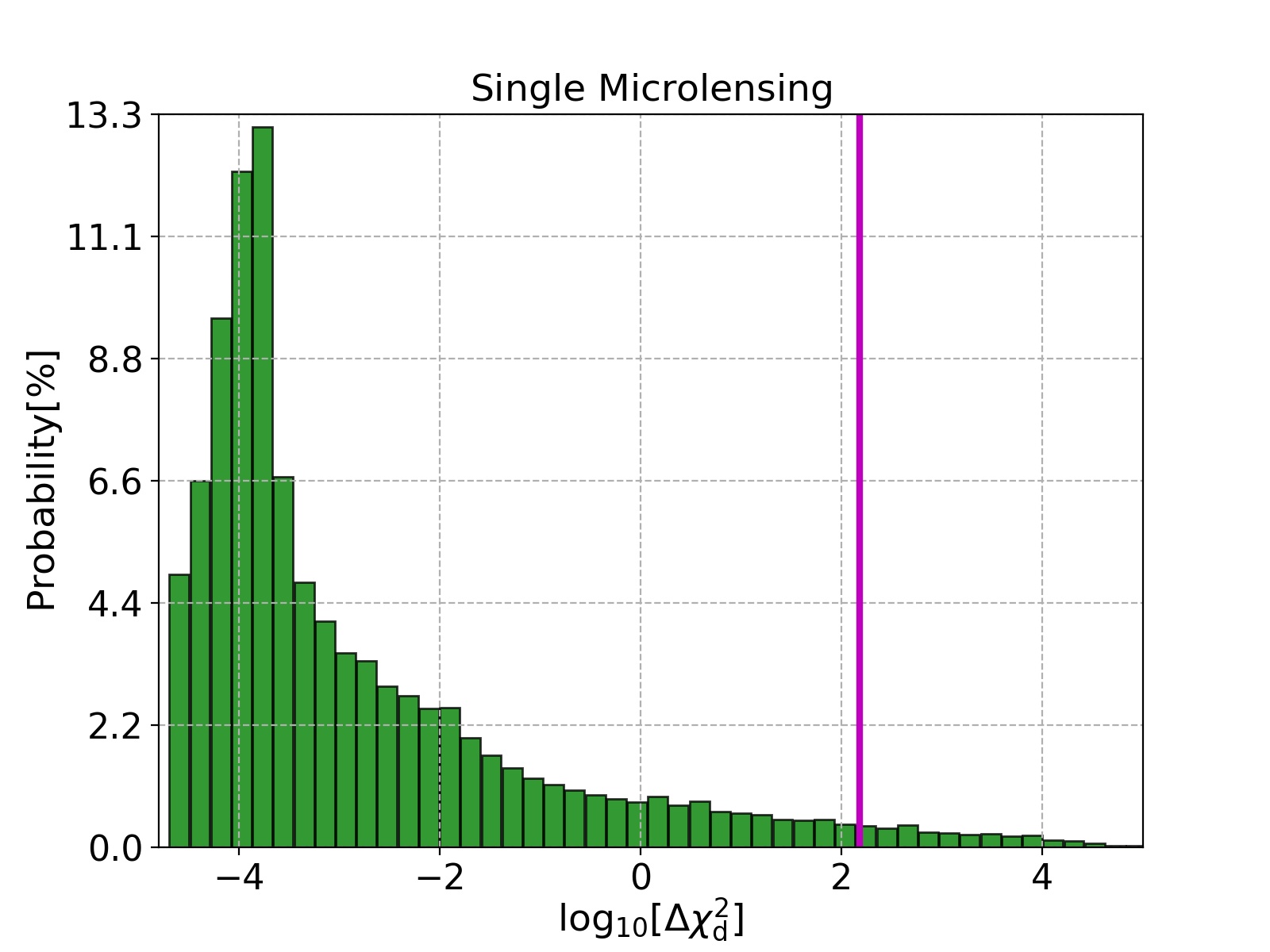}
\includegraphics[angle=0,width=0.49\textwidth,clip=0]{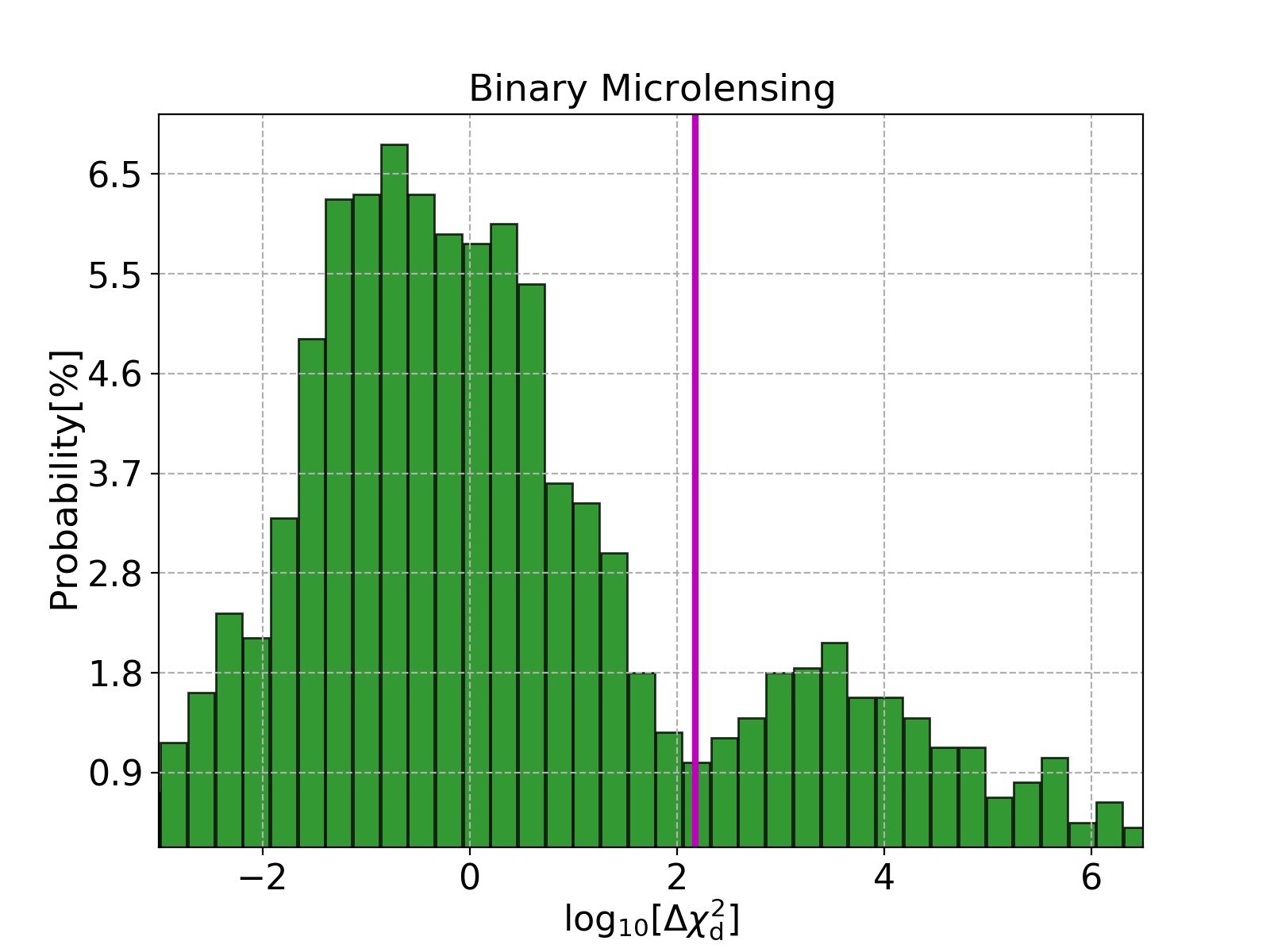}
\caption{The histograms of the $\Delta \chi^{2}_{\rm d}$ values for the simulated single (left panel) and binary (right panel) microlensing events which are detectable by \wfirst.~The vertical magenta lines specify the threshold value for detecting the disc signatures, i.e., $\Delta \chi^{2}_{\rm d}=150$.}\label{histo}
\end{figure*}

From simulated microlensing events, we extract the events which are discernible by the \wfirst~survey. In this regard, we consider three criteria which are: (i) the apparent magnitude of source star in W149 at the baseline, $m_{\rm{b}}$, should be in the range of $14.8-26~$mag, (ii) four consecutive data points should locate above the baseline by the difference larger than $4\delta_{\rm{m}}$, and (iii) $\Delta \chi^{2}_{\rm l}=|\chi^{2}_{\textrm{real}}-\chi^{2}_{\textrm{b}}|>250$, i.e., the difference of $\chi^{2}$ from fitting the real model and the baseline model should be larger than the threshold value. In the baseline model the apparent magnitude of the source star, $m_{\rm b}$, fixes versus time. In the simulation, we found that the second criterion separates the events with $\Delta \chi^{2}_{\rm l}>150$. Although,  in the third criterion we demand the larger value for $\Delta \chi^{2}_{\rm l}$ for selecting the microlensing events, the second criterion is still needed. Because it rejects the events that their high values of $\Delta \chi^{2}_{\rm l}$ are due to either a very high number of data points close to the baseline or largely sparse data on both sides of the baseline.\\

In order to indicate whether the discs around the source stars, in the confirmed microlensing events, can be distinguished we put on another criterion which is $\Delta \chi^{2}_{\rm{d}}=| \chi^{2}_{\textrm{real}}-\chi^{2}_{\textrm{best}}|>150$, i.e., the difference between the $\chi^{2}$ from fitting the real model with the disc around the source star and the best-fitted model without discs should be larger than the threshold amount of $150$. We find the best-fitted model without discs to the synthetic data points using Markov-Chain Monte-Carlo (MCMC) simulation. We use the optimization routine \texttt{EMCEE} \footnote{\url{https://emcee.readthedocs.io/en/stable/}} to make the ensemble sampler \citep{emcee,emcee2}. For single and binary microlensing light curves we consider $u_{0},~t_{0},~t_{\rm E},~\rho_{\star},~b,~m_{\rm b}$ and $u_{0},~t_{0},~t_{\rm E},~\rho_{\star},~b,~m_{\rm b},~q,~d,~\alpha$ as free parameters while fitting, respectively. Here, $\alpha$ is the angle of the source trajectory with respect to the horizontal axis. Since the disc-induced perturbations do not significantly deform the microlensing light curves, we expect best-fitted models to be very close to the real models. Hence, we start MCMC simulations from real models without discs.  

\noindent Briefly, we first extract the microlensing events detectable by \wfirst~which pass the mentioned criteria in (i), (ii) and (iii).  For these events, we then check if their disc-induced perturbations are discernible by evaluating the values of $\Delta \chi^{2}_{\rm d}$.\\

Four examples of single and binary microlensing light curves in the W149 filter for the source stars surrounded by discs are shown in Figure \ref{fig4}. In these plots, the black filled circles are the synthetic data points taken by \wfirst~in the W149 filter. The cyan curves are the best-fitted microlensing light curves. The red solid curves show the magnified apparent magnitude of the source star without discs versus time as 
\begin{eqnarray}\label{mstar}
m_{\star}=m_{\rm{b}}-2.5 \log_{10}[b~A_{\star} +1-b].
\end{eqnarray}

\noindent The blue solid curves represent the magnified apparent magnitude of the source star and its disc versus time which is given by Equation \ref{mw149}. The residuals shown with blue curves are the relative difference in the magnitude, i.e., $\delta m_{\rm{W149}}/m_{\star}= (m_{\rm{W149}}-m_{\star})/m_{\star}$ in per cent. The cyan residuals are the relative differences in the magnitude with respect to the best-fitted models. The parameters used to generate each light curve and the values of $\Delta \chi^{2}_{\rm{d}}$ are reported in Table \ref{table}. In all of these light curves, the disc-induced signatures can be discerned.\\

The histograms of $\Delta \chi^{2}_{\rm d}$ for single (left panel) and binary (right panel) microlensing events detectable by \wfirst~are shown in Figure \ref{histo}. The vertical magenta lines show the threshold amount of $\Delta \chi^{2}_{\rm d}=150$ as detectability threshold for disc signatures. For binary microlensing events, the histogram of $\Delta \chi^{2}_{\rm d}$ has two peaks that are related to the events with and without caustic-crossing features. In binary microlensing, the disc signatures are highlighted in the caustic-crossing features which results large values of $\Delta \chi^{2}_{\rm d}$ (the second smaller peak in the right-handed histogram). Totally, in $3$ and $20$ per cent of single and binary microlensing events the disc signals could be distinguished with the $\Delta \chi^{2}_{\rm{d}}>150$, respectively.\\

\begin{figure*}
	\centering
	\subfigure[]{\includegraphics[angle=0,width=0.49\textwidth,clip=0]{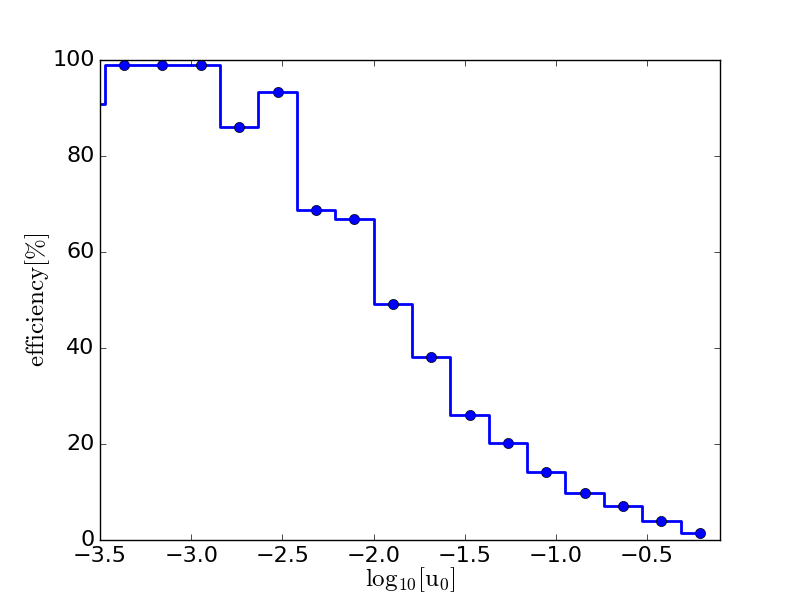}\label{fig5a}}
	\subfigure[]{\includegraphics[angle=0,width=0.49\textwidth,clip=0]{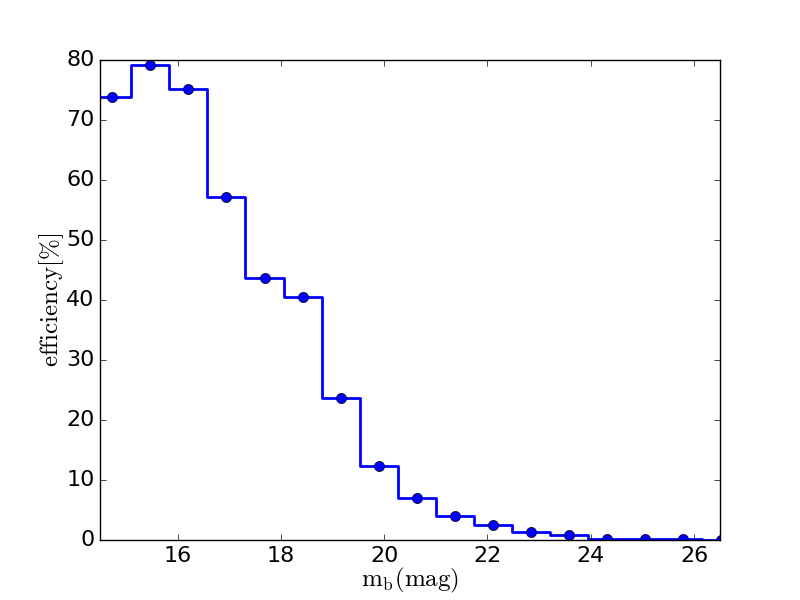}\label{fig5b}}
	\subfigure[]{\includegraphics[angle=0,width=0.49\textwidth,clip=0]{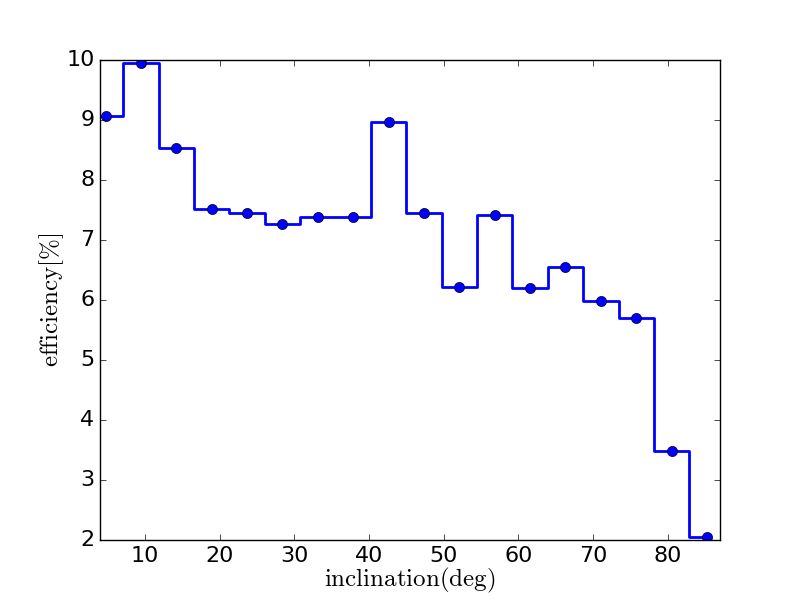}\label{fig5c}}
	\subfigure[]{\includegraphics[angle=0,width=0.49\textwidth,clip=0]{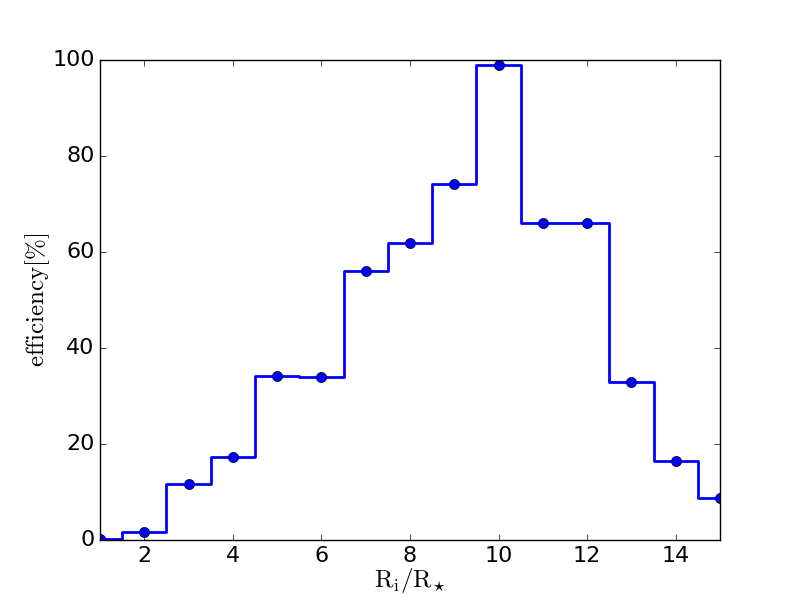}\label{fig5d}}
	\caption{The efficiencies (probability in per cent) for detecting the disc signatures in microlensing light curves versus four relevant parameters are plotted in different panels. The parameters are the lens impact parameter, the apparent magnitude of the source star in the W149 filter at the baseline, the disc inclination angle and the inner radius of disc normalized to the source radius, respectively.}\label{figf}
\end{figure*}
In order to estimate the number of discs that can be detected during \wfirst~mission, we use the results in \citet{Penny2019}. Accordingly, \wfirst~ will detect $\sim27000$ microlensing events during its $5-$year mission. We assume (i) only $\sim5$ per cent of these source stars have circumstellar discs with inner regions and (ii) around $30\%$ of stars which act as microlenses are binary and $70\%$ of them are single. We note that some of the simulated light curves resulted from binary microlenses are not distinguishable from the light curves of the single microlensing events. In real observations, the microlensing survey groups have also reported that $\sim6\%$ of the observed microlensing light curves had binarity features \citep{sumi2013, ogle2019}. 

\noindent Accordingly, \wfirst~potentially reveals the disc-induced signatures in around $28$ single microlensing events and $81$ binary microlensing ones. In these estimations, we assumed that the disc perturbations would not be misinterpreted with other kinds of perturbations. Binary microlensing events with caustic-crossing features are suitable channels for discerning the disc perturbations in microlensing surveys. This high number of detectable disc-induced signatures during \wfirst~mission emphasizes the importance of considering discs in addition to exoplanets while interpreting perturbations in microlensing light curves.\\

In order to study in what events these disc-induced signals can be detected with more probability, we plot the detection efficiency versus four relevant parameters in Figure \ref{figf} in the different panels. Chosen parameters are the lens impact parameter, the baseline source apparent magnitude in the W149 filter, the disc inclination angle and the disc inner radius normalized to the source radius. Here we summarize four remarkable points from these plots.

\begin{enumerate}[label= \arabic*., leftmargin=0.1cm]  

\item According to Figure~\ref{fig5a} the detection efficiency is a decreasing function versus the lens impact parameter. Almost all disc perturbations are detectable in high magnification microlensing events. In these events, the errors in the magnified magnitude of source stars decrease significantly. In addition, the probability of transiting inner discs by the lens is high. According to the light curves shown in Figure \ref{light}, the remarkable disc-induced deviations in the microlensing light curves occur when the lens is crossing the inner disc. \\

\item When the source star is very bright, the error in its magnitude decreases significantly (linearly in the logarithmic scale). It causes the disc-induced perturbations in the magnification factor to be highlighted. As a result, the disc detection efficiency reduces with increasing the source apparent magnitude, as shown in Figure \ref{fig5b}. However, the probability of detecting the disc perturbations for very bright source stars slightly drops. Because the stars with $m_{\rm{b}}<15~$mag are most likely hot and the edge of inner disc (where the temperature is $\sim2500~$K) gets away from the source stars. In that case, the probability of crossing this inner edge by the lens decreases. \\

\item According to Figure \ref{fig5c} the larger disc inclination angle, the relatively lower efficiency for detecting disc signatures. However, the efficiency does not strongly depend on the disc inclination angle. Two reasons can be (i) the disc-induced perturbations for high inclined discs are relatively shorter than those made by low inclined discs and (ii) crossing the inner disc of high inclined discs is not possible when the lens is passing almost parallel with the disc semi-major axis.\\

\item In the last panel (Figure \ref{fig5d}), the detection efficiency is shown versus the inner radius $R_{\rm i}$ which is normalized to the host star radius. The position of the $R_{\rm i}$ depends on the source temperature. The larger $R_{\rm i}$ means the host star is hotter and most likely brighter with the smaller error in its magnitude. The smaller error causes the weaker perturbations to be detectable. However, for very large inner radii, the efficiency drops significantly. In these cases, the probability of crossing the disc inner radius by the lens decreases when the magnification is high. \\

\end{enumerate}

\noindent We note that the disc mass and as a result $\tau_{\rm{sc}}$ are irrelevant parameters to the detection efficiency in W149. Because the disc mass affects on the reflected emission from the outer disc, but not on the thermal emission from the inner disc.

\section{conclusions}\label{five}

\wfirst~ will be launched in near future and probe microlensing events towards the Galactic bulge in six $72$-day seasons during its $5$-year mission \citep{spergel2015}. It observes the Galactic bulge with the short cadence, $15.16~$min, in the wide near-infrared filter W149 \footnote{\url{https://wfirst.ipac.caltech.edu/sims/Param_db.html}}. Accordingly, we expect that the anomalies of source stars which have near-infrared radiation are magnified or event detected through microlensing observations. One possible origin of these extra radiations is the inner region of circumstellar discs around the source stars. This subject was studied in this work.\\

\wfirst~will observe in the W149 filter which its mean wavelength, $1.49~\mu$m, coincides with the maximum thermal emission from inner discs. On the other hand, source stars of microlensing events in \wfirst~survey are, on average, faint and most likely cool. The discs around these stars survive longer than those around hotter ones \citep{Pascucci2011}. Hence, the discs can be found with higher probability around the source stars of \wfirst~microlensing events. \\

In single microlensing events, discs generally break the symmetry of simple light curves with respect to the time of closest approach. If the disc is facing on or the lens trajectory is parallel with (or normal to) the semi-major axis of the projected disc on the sky plane, the resulted light curve is symmetric. Discs generate one or two extra peak(s) at the time of closest approach to the disc inner radius or when the lens is crossing it. The thermal emission from the inner disc maximizes in the near-infrared and at wavelengths of some micrometers. Hence, it is mostly detectable in the W149 filter. The reflected radiation from the outer discs maximizes in the visible band and can be detected in the Z087 filter.\\

In caustic-crossing binary microlensing of source stars surrounded by discs, two extra extended peaks form right before entering and immediately after exiting from caustic curves. The discs cause that the magnification peak in caustic-crossing features reduces, similar to the finite size effect. If the disc size is on the order of caustic size, the disc slightly decreases the magnification factor.\\

By doing a Monte-Carlo simulation, we deduced that the efficiencies for detecting the disc-induced perturbations in the W149 filter in single and binary microlensing events are $3$ and $20$ per cent, respectively. If we assume that (i) $\sim5$ per cent of the source stars have discs with inner regions and (ii) around $30$ per cent of microlenses are binary systems, the disc signatures can be distinguished with $\Delta \chi^{2}_{\rm{d}}>150$ in $28$ and $81$ single and binary microlensing light curves detected by \wfirst,~respectively. Totally, $109$ detectable disc-induced signatures in \wfirst~light curves highlight the importance of considering discs in addition to exoplanets while interpreting anomalies in light curves as generators of perturbations. \\

\section*{Data availability}
Data available on request.

\section*{Acknowledgements}
We especially acknowledge C.~Han for reading and commenting on the paper. We also thank R.~Poleski for consultation and the anonymous referee for his/her careful comments. 

\bibliographystyle{mnras}
\bibliography{reference}
\end{document}